\documentclass[useAMS,usenatbib,usegraphicx]{mn2e}

\usepackage{natbib}
\usepackage{amsmath}
\usepackage{txfonts}
\usepackage{color}
\usepackage{array}
\usepackage{graphicx}
\usepackage{epstopdf}
\usepackage{color}
\usepackage{bm}
\usepackage{cmap}
\usepackage{booktabs}
\usepackage{tabularx}
\usepackage{multicol}
\usepackage{float}

\DeclareMathOperator\erf{erf}

\begin{document}
\title[Statistical theory of thermal evolution of neutron stars]{
Statistical theory of thermal evolution of neutron stars}

\author[M. V. Beznogov and D. G. Yakovlev]{
M. V. Beznogov$^{1}$\thanks{E-mail: mikavb89@gmail.com},
D. G. Yakovlev$^{2}$
\\
$^{1}$St.~Petersburg Academic University, 8/3 Khlopina st.,
St.~Petersburg 194021, Russia \\
$^{2}$Ioffe Physical Technical Institute, 26 Politekhnicheskaya st.,
St.~Petersburg 194021, Russia}

\date{Accepted . Received ; in original form}
\pagerange{\pageref{firstpage}--\pageref{lastpage}} \pubyear{2014}
\maketitle \label{firstpage}

\begin{abstract}
Thermal evolution of neutron stars is known to depend on the
properties of superdense matter in neutron star cores. We suggest
a statistical analysis of isolated cooling middle-aged neutron stars
and old transiently accreting quasi-stationary neutron stars warmed
up by deep crustal heating in low-mass X-ray binaries. The method is
based on simulations of the evolution of stars of different masses
and on averaging the results over respective mass distributions.
This gives theoretical distributions of isolated neutron stars in
the surface temperature--age plane and of accreting stars in the
photon thermal luminosity--mean mass accretion rate plane to be
compared with observations. This approach permits to explore not
only superdense matter but also the mass distributions of isolated
and accreting neutron stars. We show that the observations of these
stars can be reasonably well explained by assuming the presence of
the powerful direct Urca process of neutrino emission in the inner
cores of massive stars, introducing a slight broadening of the
direct Urca threshold (for instance, by proton superfluidity), and
by tuning mass distributions of isolated and accreted neutron stars.
\end{abstract}


\section{Introduction} \label{sec:intro}

In this paper we study neutron stars of two types. First, they are
\emph{cooling isolated middle-aged} ($10^2-10^6$ yr) neutron stars
which are born hot in supernova explosions but gradually cool down
mostly via neutrino emission from their superdense cores. They are
mainly thermally relaxed and isothermal inside. A noticeable
temperature gradient still persists only in their thin heat
blanketing envelopes (e.g., \citealt*{GPE83,Potekhin_etal97}).

Secondly, we study old ($t \gtrsim 10^8-10^9$ yr) \emph{transiently
accreting quasi-stationary neutron stars in low-mass X-binaries}
(LMXBs); such transient systems are called X-ray transients (XRTs).
These neutron stars accrete matter from time to time (in the active
states of XRTs) from their low-mass companions. The accreted matter
is compressed under the weight of newly accreted material and the
compression is accompanied by deep crustal heating
\citep*{HZ90,HZ08} due to beta-captures, neutron absorption and
emission, and pycnonuclear reactions with characteristic energy
release of 1--2 MeV per accreted nucleon deeply in the crust. The
accretion episodes are supposed to be neither too long (months--
weeks) nor too intense to overheat the crust and destroy the
internal equilibrium between the crust and the core. Nevertheless,
the deep crustal heating should be sufficiently strong to keep the
neutron stars warm and explain observable thermal emission of such
neutron stars during quiescent states of XRTs \citep*{BBR98}. The
mean neutron star heating rate is determined by the average mass
accretion rate $\langle \dot{M} \rangle$; the averaging has to be
performed over characteristic cooling times of such stars (typically
$\gtrsim 10^3$ yr).

The isolated cooling neutron stars are usually studied by
calculating their \emph{theoretical cooling curves} (time dependence
of their effective surface temperature $T_{\rm s}(t)$ or
(equivalently) thermal surface luminosity $L_\gamma(t)$, redshifted
or non-redshifted for a distant observer). The curves are calculated
under different assumptions on the neutrino emission in the stellar
core, and then they are compared with observations (to reach the
best agreement).

The transiently accreting neutron stars in XRTs are investigated by
simulating their \emph{theoretical heating curves}, which give
average $T_{\rm s}$ or $L_\gamma$ for accreting neutron stars in
quiescent states as a function of $\langle \dot{M} \rangle$. The
heating curves are also compared with observations.

It is important that the cooling and heating curves have much in
common (e.g., \citealt*{YH03,YLH03}) and allow one to study
fundamental physics of neutron stars. As a rule, one plots the
cooling and heating curves to interpret observations of individual
stars. The most important cooling/heating regulators to be tested
are as follows.
\begin{enumerate}

\item
A level of neutrino luminosity of the star. Specifically, it is the
neutrino cooling rate $L_\nu/C$ for a cooling neutron star or
neutrino luminosity $L_\nu$ for a transiently accreting star ($C$
being the heat capacity of the star).

\item
Stellar mass and equation of state (EOS) of superdense matter in the
stellar core which regulate the level of the neutrino emission in
the core.

\item
Composition of the heat blanketing envelope of a cooling or heating
star which determines the relation between the internal and surface
temperature of the star.

\end{enumerate}

Since observations of isolated and transiently accreting neutron
stars are rapidly progressing, it is instructive to utilize the
accumulated statistics of the sources and develop a statistical
theory of their evolution. It is the aim of this paper to put
forward such a theory. It will take into account that the cooling
and heating curves can strongly depend on neutron star masses.
Then, one can introduce the probability to find a source in different
places of the cooling/heating diagram by averaging these curves over
mass distributions of isolated or accreting stars. Naturally, these
mass distributions can be different. Comparing theoretical and
observational distributions of the sources one can study not only
individual cooling regulators mentioned above but also the mass
distributions of neutron stars of different types. The problem would
be to find out which physical models of neutron stars and mass
distributions of isolated {\it and} accreting neutron stars give the
best agreement of calculated and observed distributions of such
stars on the cooling {\it and} heating diagrams.

\section{Observational basis}
\label{s:observa}

\begin{figure}
\centering
\includegraphics[width=0.51\textwidth]{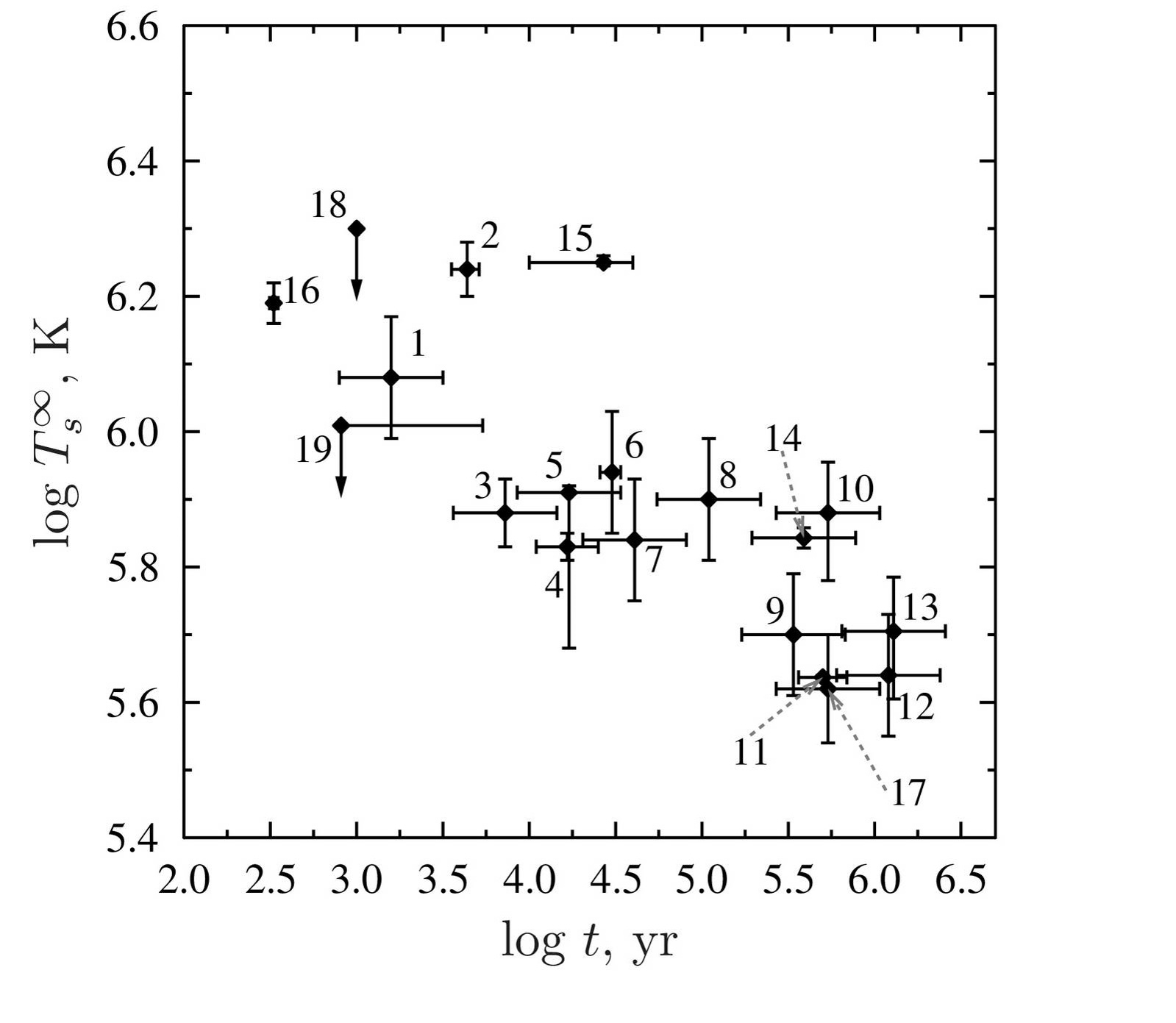}
\caption{Logarithms of effective surface
temperatures and ages of cooling isolated middle-aged neutron stars
which show thermal emission from their surfaces (inferred or
constrained from observations). The source numbers are the
same as in Table \ref{tab:cool}.}
\label{fig:cool}
\end{figure}

\begin{figure}
\centering
\includegraphics[width=0.51\textwidth]{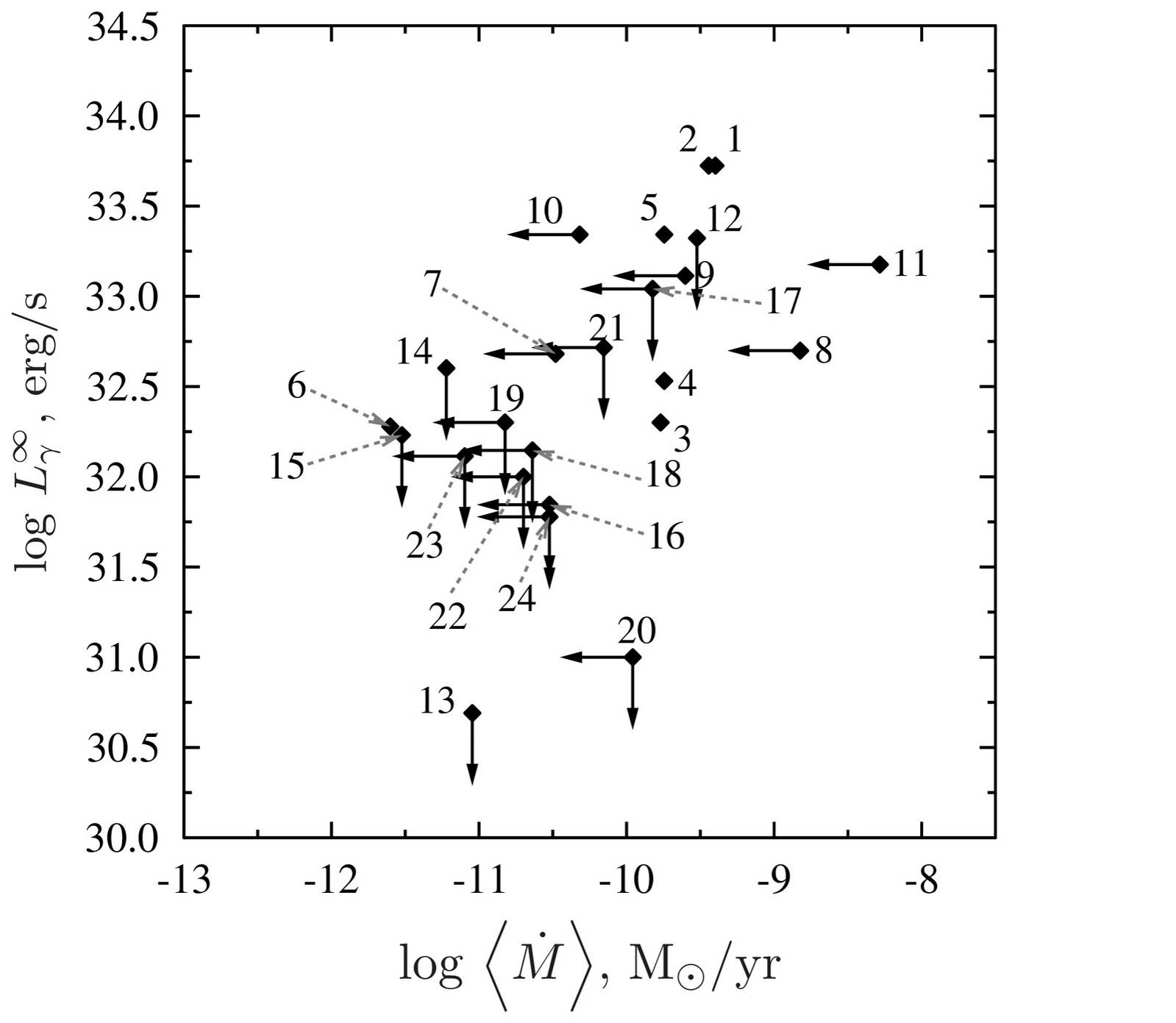}
\caption{Logarithms of surface thermal quiescent
luminosities $L_\gamma^\infty$ (redshifted for a distant observer)
and mean mass accretion rates $\langle \dot{M} \rangle$ of
transiently accreting neutron stars in SXTs (inferred or
constrained from observations). Numeration of the sources is the
same as in Table \ref{tab:heat}.}
\label{fig:heat}
\end{figure}

Before describing statistical theory of thermal evolution of neutron
stars, in Tables \ref{tab:cool} and \ref{tab:heat} and Figs
\ref{fig:cool} and \ref{fig:heat} we present the observational basis
for our analysis.

Table \ref{tab:cool} gives the data on 19 isolated middle-aged
neutron stars whose thermal surface radiation has been detected or
constrained. The table gives the source number, source name,
estimated age, the effective surface temperature $T_{\rm s}^\infty$
(redshifted for a distant observer) as inferred from observations,
the confidence level of $T_{\rm s}^\infty$, a model which has been
used to infer $T_{\rm s}^\infty$, and references to original
publications from which the results are taken. The data have been
collected in the same way as in \citet*{YP04} and \citet{YGKP08} but
supplemented by new results. The data contain neutron stars in
supernova remnants (like the Crab pulsar), the famous Vela pulsar
and its twin PSR 1706--44, compact stellar objects in supernova remnants (like neutron star in Cas A), the ``dim'' (``truly''
isolated) stars (e.g., RX J1865.4--3754), etc. The distances are not
very certain even if parallaxes are measured (see a discussion on RX
J1865.4--3754 in \citealt{Potekhin_14}). In many cases the ages are
uncertain as well. The presented values of $T_{\rm s}^\infty$ refer
to thermal emission from the entire surface of the stars. These
temperatures are inferred from the observed spectra using blackbody
(BB) model for thermal emission, the models of nonmagnetic and
magnetic hydrogen atmospheres (HA and mHa, respectively), the models
of hydrogen atmospheres of finite depth, HA$^*$ and mHA$^*$, as well
as the carbon atmosphere (CA) models (as reviewed, e.g., by
\citealt{Potekhin_14}).

Table \ref{tab:heat} gives the data on neutron stars in 26 XRTs. It
presents the number and name of the source, estimated (constrained)
mean mass accretion rates $\langle \dot{M} \rangle$, themal surface
luminosities of neutron stars $L_\gamma^\infty$ in quasi-stationary
quiescent states, and respective references. Extracting  $\langle
\dot{M} \rangle$ and  $L_\gamma^\infty$ from observations is a very
complicated problem as discussed in many references cited in Table
\ref{tab:heat}. Both quantities are often constrained (given as
upper limits) rather than measured. If measured, their values are
rather uncertain (the error bars are large and difficult to
estimate, often not presented). Therefore, one should be careful in
dealing with these data. The statistical approach we describe here
seems most suitable for this situation.

\section{Statistical theory}
\label{s:stat-theory}

Now we present the simplest version of the statistical theory for
cooling isolated neutron stars and heating transiently accreting
quasi-stationary neutron stars in LMXBs.

\begin{table*}
\centering \caption{Middle-aged cooling isolated neutron stars whose
thermal surface emission has been detected or constrained; see text
for details}
    \begin{tabular}{c l c c c l l}
    \toprule
    Num.  &  Source  &  Age, kyr  &  $T_s^\infty $, MK  &  Confid.\ level for $T_s^\infty $  &  Model  &  Ref. \\ 
    \midrule
    1  &  PSR J1119--6127  &  $\sim$1.6   &  $\approx$1.2  &  --  &  mHA  & Z09  
 \\
    2  &  RX J0822--4300 (in Pup A)  &  $4.4 \pm 0.8$  &  1.6--1.9  &  90\%  &  HA  &  Z99,
    B12 
 \\
    3  &  PSR J1357--6429  & $\sim 7.3$  &  $\approx 0.77$  &  --  &  mHA  & Z07 
 \\
    4  &  PSR B0833--45 (Vela)  &  11--25  &  $0.68\pm 0.03$  &  68\%  &  mHA  & P01 
  \\
    5  &  PSR B1706--44  &  $\sim$17  &  $0.82^{+0.01}_{-0.34}$  &  68\%  &  mHA  & MG04
\\
    6  &  PSR J0538+2817  &  $30\pm 4$   &  $\sim 0.87$  &  --  &  mHA  & Z04 
\\
    7  & PSR B2334+61  &  $\sim$41  &  $\sim 0.69$  &  --  &  mHA  &  Z09 
\\
    8  &  PSR B0656+14  &  $\sim$110  &  $\sim 0.79$  &  --  &  BB  & Z09 
\\
    9  & PSR B0633+1748 (Geminga)  &  $\sim$340  &  $0.5\pm 0.1$  &  --  &  BB  &  K05
\\
    10  & PSR B1055--52  &  $\sim$540  &  $\sim 0.75$  &  --  &  BB  & PZ03 
\\
    11  &  RX J1856.4--3754  &  $\sim$500  &  $0.434 \pm 0.003$  &  68\%  &  mHA$^*$  & Ho07, P14 
\\
    12  & PSR J2043+2740  &  $\sim$1200   &  $\sim 0.44$  &  --  &  mHA  & Z09  
\\
    13  & RX J0720.4--3125  &  $\sim$1300  &  $\sim 0.51$  &  --  &  HA$^*$  & M03  
\\
    14  & PSR J1741--2054   &  $\sim$391  &  $0.70\pm 0.02$  &  90\%  &  BB  & Ka14 
\\
    15  & XMMU J1732--3445  &  $\sim$27   &  $1.78^{+0.04}_{-0.02}$   &  --  &  CA  & K15 
\\
    16  & Cas A NS &  $0.33$   &  $\approx 1.6$  &  --  &  CA  & H09 
 \\
    17  &  PSR J0357+3205  (Morla) & $\sim$540   &  $0.42^{+0.09}_{-0.07}$  & 90\%  &  mHA  & M13, Ki14 
\\
    18  &  PSR B0531+21 (Crab)  &  1  &  $<2.0$  &  99.8\%  &  BB  & W04, W11
\\
    19  &  PSR J0205+6449 (in 3C 58)  &  0.82--5.4  &  $<1.02$  &  99.8\%  &  BB  &  S04, S08 
\\
    \hline
  \end{tabular}
      \begin{tabular}{lp{16cm}}
& [Z09] \citet*{Zavlin09};
    [Z99] \citet*{ZTP99};
    [B12] \citet{BPWP12};
    [Z07] \citet*{Zavlin07};
    [P01] \citet{Pavlov_etal01};
    [MG04] \citet{McGowan_etal04};
    [Z04] \citet*{ZP04};
    [K05] \citet{KPZR05};
    [PZ03] \citet*{PZ03};
    [Ho07] \citet{Ho_etal07};
    [P14] \citet*{Potekhin_14};
    [M03] \citet*{MZH03};
    [Ka14] \citet{Karpova_etal14};
    [K15] \citet{Klochkov_etal15};
    [H09] \citet*{HoHeinke_09};
    [M13] \citet{Marelli_etal13};
    [Ki14] \citet{Kirichenko_etal14};
    [W04] \citet{Weisskopf_etal04};
    [W11] \citet{Weisskopf_etal11};
    [S04] \citet{SHSM04};
    [S08] \citet{Shibanov_etal08}.
    \end{tabular}
\label{tab:cool}
\end{table*}

\begin{table*}
\centering \caption{Accreting neutron stars in XRTs whose surface
thermal emission in quasi-stationary quiescent state has been
detected or constrained;  see text for details.}
    \begin{tabular}{c l c c l}
    \toprule
    Num.  &  Source  &     $\dot{M}$, $\mathrm{M}_\odot$ yr$^{-1}$  &  $L_\gamma^\infty$, erg~s$^{-1}$  &  Ref. \\ 
    \midrule
    1     &  Aql X-1  &   $4\!\times\!10^{-10}$ & $5.3\!\times\!10^{33}$  &  H07, R01a, C03, T04
\\
    2     &  4U 1608--522  &    $3.6\!\times\!10^{-10}$  &  $5.3\!\times\!10^{33}$  &  H07, T04, R99
\\
    3     &  MXB 1659--29  &    $\!1.7\times\!10^{-10}$  &  $2.0\!\times\!10^{32}$  &  H07, C06a
\\
    4     &  NGC 6440 X-1  &   $1.8\!\times\!10^{-10}$  &  $3.4\!\times\!10^{32}$  & H07, C05
\\
    5   &  RX J1709--2639  &    $1.8\!\times\!10^{-10}$  & $2.2\!\times\!10^{33}$  & H07, J04a
\\
    6   &  IGR 00291+5934  &    $2.5\!\times\!10^{-12}$  &  $1.9\!\times\!10^{32}$  & H09b, G05, J05, T08
\\
    7     &  Cen X-4  &    $<3.3\!\times\!10^{-11}$  &  $4.8\!\times\!10^{32}$  &  T04, R01b
\\
    8     &  KS 1731--260  &    $<1.5\!\times\!10^{-9}$  &  $5\!\times\!10^{32}$  & H07, C06a
\\
    9   &  1M 1716--315  &    $<2.5\!\times\!10^{-10}$  &  $1.3\!\times\!10^{33}$  &
J07a, H09b
\\
    10  &  4U 1730--22  &  $<4.8\!\times\!10^{-11}$  &  $2.2\!\times\!10^{33}$  & H09b, T07, C97
\\
    11  &  4U 2129+47    &  $<5.2\!\times\!10^{-9}$  &  $1.5\!\times\!10^{33}$  & H09b, N02, P86, W83
\\
    12   &  Terzan 5    &  $3\!\times\!10^{-10}$  &  $<2.1\!\times\!10^{33}$  & H07, H06b,
    W05a
\\
    13   &  SAX J1808.4--3658    &  $9\!\times\!10^{-12}$  & $<4.9\!\times\!10^{30}$  & H09b, GC06, CS05
\\
    14   &  XTE J1751--305    &  $6\!\times\!10^{-12}$  &  $<4\!\times\!10^{32}$  &  H09b, M02, M03, W05b
\\
    15   &  XTE J1814--338  &  $3\!\times\!10^{-12}$  &  $<1.7\!\times\!10^{32}$  &
    H09b, K05, W03, G06
\\
    16     &  EXO 1747--214 &  $<3\!\times\!10^{-11}$  &  $<7\!\times\!10^{31}$  &
    T05, H07
\\
    17     &  Terzan 1  &  $<1.5\!\times\!10^{-10}$  & $<1.1\!\times\!10^{33}$  &
  C06b, H07 \\
    18   &  XTE J2123--058   &  $<2.3\!\times\!10^{-11}$  &  $<1.4\!\times\!10^{32}$   &
    H07, T04 
\\
    19   &  SAX J1810.8--2609  &  $<1.5\!\times\!10^{-11}$  &  $<2.0\!\times\!10^{32}$  &
    H07, T04, J04b 
\\
    20   &  1H 1905+000   &  $<1.1\!\times\!10^{-10}$  & $<1.0\!\times\!10^{31}$  &
    J06, J07b, H09b    %
\\
    21   &  2S 1803--45   &  $<7\!\times\!10^{-11}$  &  $<5.2\!\times\!10^{32}$
    & H09b, C07 
\\
    22   &  XTE J0929--314    &  $<2.0\!\times\!10^{-11}$  & $<1.0\!\times\!10^{32}$  &
    G02, G06, J03, CF05, W05b, H09b
\\
    23   &  XTE J1807--294   &  $<8\!\times\!10^{-12}$  &  $<1.3\!\times\!10^{32}$  &
    H09b, G06, CF05 
\\
    24   &  NGC 6440 X-2   &  $<3\!\times\!10^{-11}$  &  $<6\!\times\!10^{31}$  &
    H10 
\\
    \hline
    \end{tabular}
    \begin{tabular}{lp{16cm}}
& [H07] \citet{Heinke07};
    [R01a] \citet{Rutledge_etal01a};
    [C03] \citet*{CS03};
    [T04] \citet{TGHK04};
    [R99] \citet{Rutledge_etal99};
    [C06a] \citet{Cackett_etal06a};
    [J04a] \citet{Jonker_etal04a};
    [H09b] \citet{Heinke09b};
    [G05] \citet{Galloway_etal05};
    [J05]  \citet{Jonker_etal05};
    [C05] \citet{Cackett_etal05};
    [T08] \citet{Torres_etal08};
    [R01b] \citet{Rutledge_etal01b};
    [J07a] \citet{JBW07a};
    [T07] \citet*{TGK07};
    [C97] \citet*{CSL97};
    [N02] \citet*{NHB02};
    [P86] \citet{PSGG86};
    [W83] \citet{Wenzel83};
    [H06b] \citet{Heinke_etal06b};
    [W05a] \citet{Wijnands_etal05};
    [GC06] \citet*{GC06};
    [CS05] \citet{Campana_etal05}
    [M02] \citet{Markwardt_etal02};
    [M03] \citet{Miller_etal03};
    [W05b] \citet{Wijnands_etal05b};
    [K05] \citet{Krauss_etal05};
    [W03] \citet*{WR03};
    [G06] \citet*{Galloway06};
    [T05] \citet*{TGK05};
    [C06b] \citet{Cackettetal06b};
    [J04b] \citet*{JWK04b};
    [J06] \cite{Jonker_etal06};
    [J07b] \citet{JSCJ07b};
    [C07] \citet*{CWH07};
    [G02] \citet{GCMR02};
    [J03] \citet*{JGC03};
    [CF05] \citet{CFSI05};
    [H10] \citet{Heinke10}
    \end{tabular}
\label{tab:heat}
\end{table*}

\renewcommand{\arraystretch}{1.2}
\begin{table}
\caption{Masses, radii, and central densities of two neutron star
models with HHJ EOS}
\begin{center}
\begin{tabular}{ c l l  l }
\toprule
Model & $M/{\rm M_\odot}$ & $R$ (km) & $\rho_{c14}$   \\
\midrule
Maximum mass & 2.16    &  10.84 & 24.5   \\
Direct Urca onset & 1.72     &  12.46 & 10.0    \\
 \hline
\end{tabular}
\label{tab:models}
\end{center}
\end{table}

The theory is based on \emph{ordinary} theory of neutron star
cooling and heating (e.g., \citealt{Page_etal09,YP04}). By way of
illustration we consider neutron stars with nucleonic cores and some
phenomenological EOS in the core described by \citet{KKPY14}. The authors denoted this EOS as HHJ; it belongs to the family of
parameterized EOSs suggested by \citet{HHJ99}. The parameters of two
HHJ models (gravitational masses $M$, circumferential radii $R$, and
central densities $\rho_{\rm c}$ in units of $10^{14}$ g~cm$^{-3}$)
are presented in Table \ref{tab:models}. The first is the maximum
mass model, with $M_{\rm max}=2.16\,{\rm M_\odot}$ (to be consistent
with recent measurements of masses $M\approx 2 {\rm M_\odot}$ of two
neutron stars by \citealt{Demorest_etal10} and
\citealt{Antoniadis_etal13}). The circumferencial radius of the most
massive stable star in this case is $R=10.84$~km and the central
density $\rho_{\rm c}= 2.45 \times 10^{15}$~g~cm$^{-3}$. The second
model in Table \ref{tab:models} is the model with $M=M_{\rm
D}=1.72\,{\rm M_\odot}$. At lower $M$ the powerful direct Urca
process of neutrino emission \citep{LPPH91} is forbidden in a
neutron star core, while at higher $M$ it is allowed in the central
kernel of a star (at densities $\rho>\rho_{\rm D}= 1.00 \times
10^{15}$~g~cm$^{-3}$). Such high-mass stars undergo very rapid
neutrino cooling.

We calculate thermal evolution of cooling and heating neutron star
models using our generally relativistic cooling code \citep{GYP01}
on a dense grid of masses $M$, from $1.1\, \mathrm{M}_\odot$ to
$2.1\,\mathrm{M}_\odot$. The cooling curves of isolated neutron stars are
obtained by directly running the code (although we are most interested
in the ages from $10^2$ to $10^6$ yr at which the stars are
isothermal inside and cool via neutrino emission so that the cooling
problem is considerably simplified).

The heating curves of transiently accreting neutron stars are
calculated as stationary solutions of the heat balance equation
(e.g., \citealt{HZ03}),
\begin{equation}
     L_{\rm h}^\infty=L_\nu^\infty+L_\gamma^\infty,
\label{e:heat}
\end{equation}
where $L_{\rm h}^\infty$ is the averaged deep crustal heating power
(redshifted for a distant observer and determined by the
time-averaged mass accretion rate $\langle \dot{M} \rangle$). The
interior of the star is assumed to be isothermal (with general
relativistic effects properly included) while the internal
temperature is related to the effective surface one by corresponding
heat blanketing solution (e.g., \citealt{Potekhin_etal97}). Calculated
cooling curves will be plotted on the $T_{\rm s}^\infty-t$ diagram,
while heating curves will be plotted on the $L_\gamma^\infty-
\langle \dot{M} \rangle$ diagram. These will be ordinary cooling and
heating curves which have been extensively studied by the theory.
As a rule, the highest cooling or
heating curve corresponds to the low-mass
neutron star (with rather slow neutrino emission) while the lowest
curve belongs to the maximum-mass star with highest neutrino cooling
rate. The space between the highest and lowest cooling curves is
filled by the curves for stars of different masses $M$ but this
filling can be very non-uniform (e.g., \citealt{GKYG05}).

%
\begin{figure}
\centering
\includegraphics[width=0.51\textwidth]{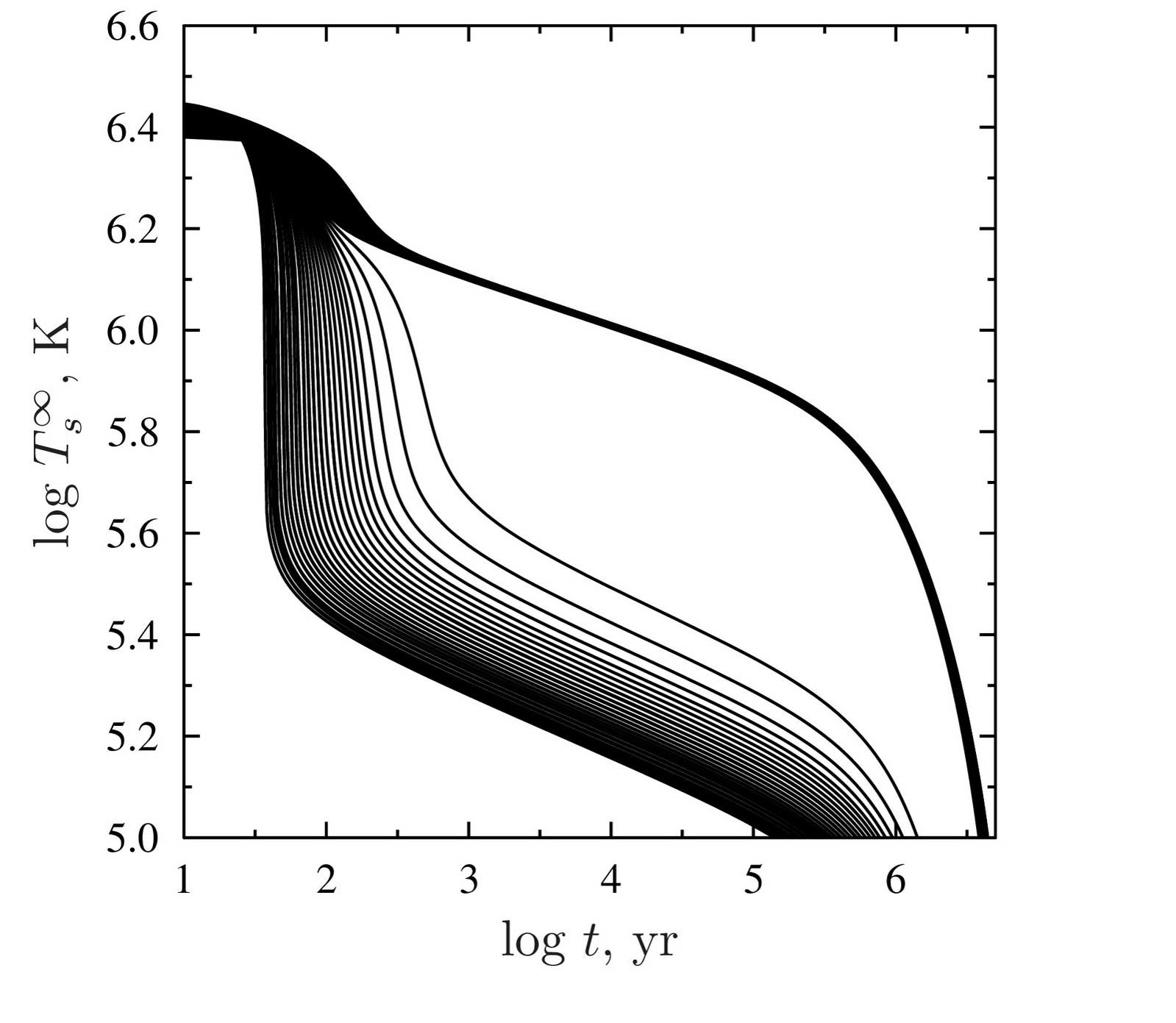}
\caption{A sequence of cooling curves $T_{\rm
s}^\infty(t)$ of neutron stars of masses $M=1.1\,\mathrm{M}_\odot-2.1\,
\mathrm{M}_\odot$ with mass difference of stars for neighboring curves $\Delta
M=0.01\,{\mathrm{M}_\odot}$. The heat blanketing envelope is made of
iron. The threshold for the onset of the direct Urca process is not
broadened (as detailed in Sections \ref{s:stat-theory} and \ref{s:broadUrca}).}
\label{fig:theorcoolNA}
\end{figure}
\begin{figure}
\includegraphics[width=0.51\textwidth]{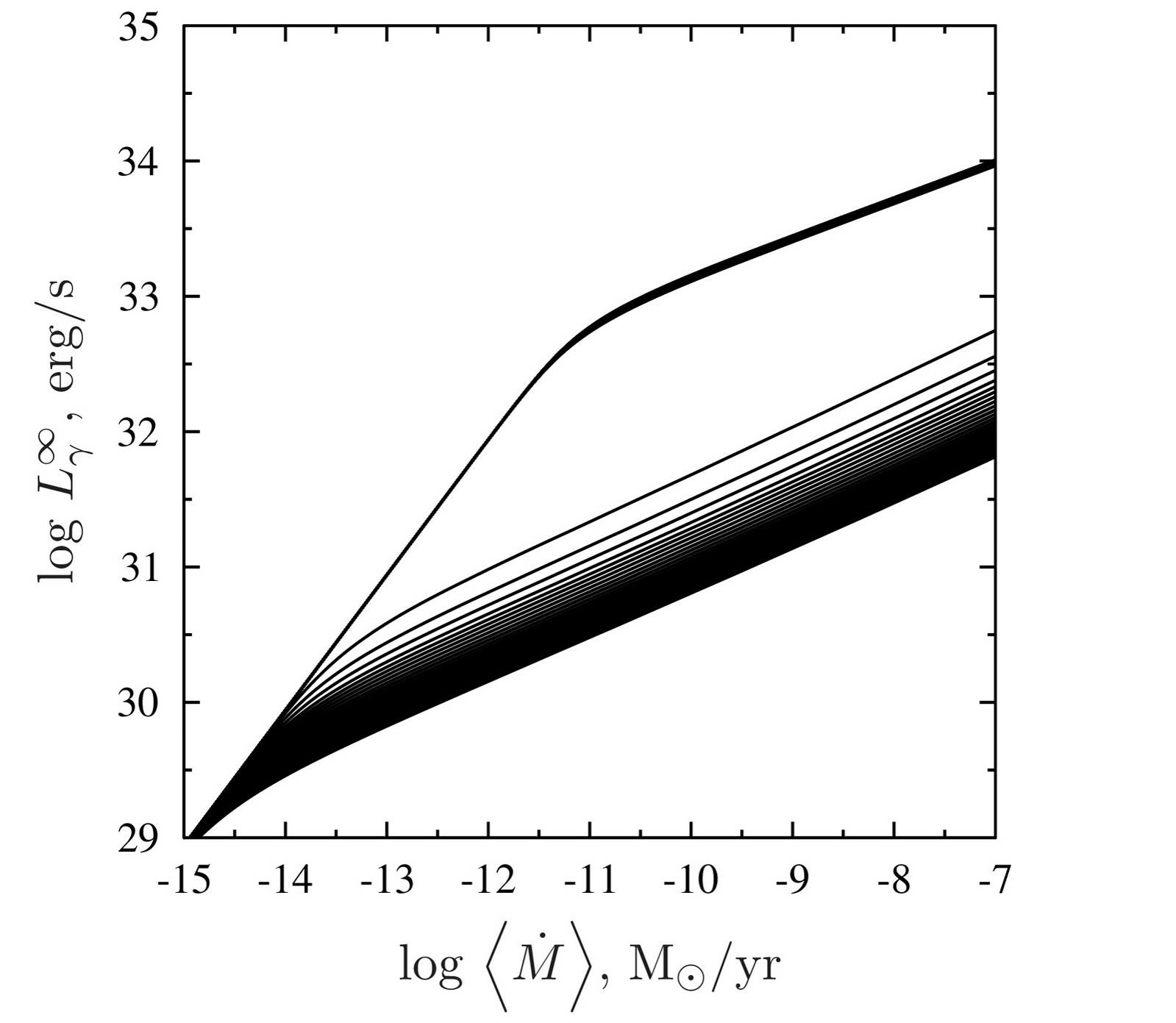}
\caption{ A sequence of heating curves $L_\gamma
^\infty\left(\langle \dot{M} \rangle\right)$ of transiently accreting neutron
stars of masses $M=1.1\,\mathrm{M}_\odot-2.1\,\mathrm{M}_\odot$, with mass
step $\Delta M=0.01\,{\mathrm{M}_\odot}$. The
heat blanketing envelope is made of iron. The direct Urca
threshold is not broadened (see text for details).}
\label{fig:theorheatNA}
\end{figure}

\begin{figure}
\centering
\includegraphics[width=0.51\textwidth]{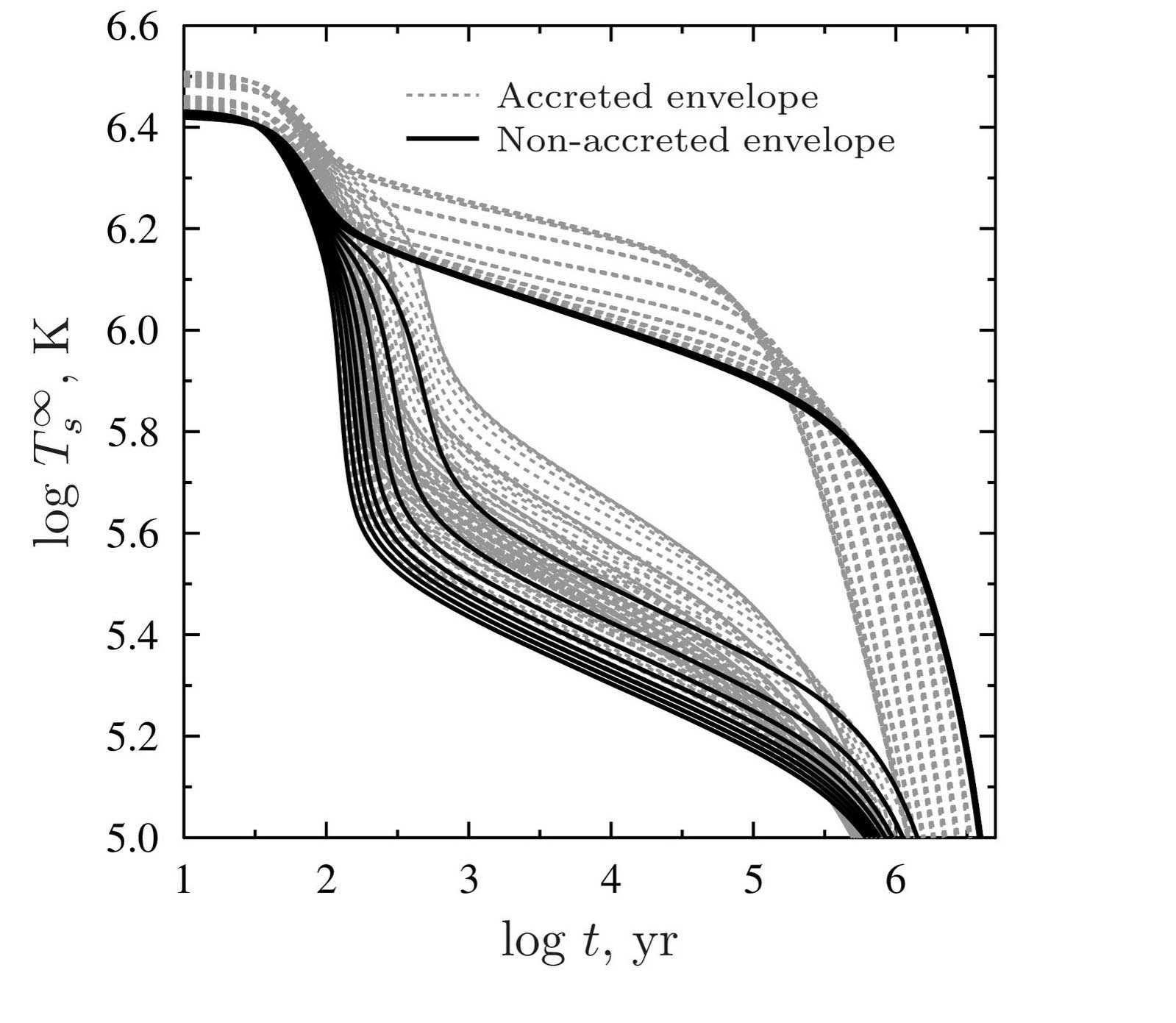}
\caption{A sequence of cooling curves $T_{\rm
s}^\infty(t)$ of neutron stars of masses
$M=1.6\,\mathrm{M}_\odot-1.8\,\mathrm{M}_\odot$
with step $\Delta M=0.01\,{\mathrm{M}_\odot}$. Solid curves correspond to iron
heat blanketing envelope, while dashed curves are for envelopes
containing light (accreted) elements
of different mass, $\Delta M_{\rm le}
=10^{-k}\,M$, where $k$=7,8, \ldots, 16. The direct Urca threshold
is not broadened (see  Sections \ref{s:stat-theory} and \ref{s:broadUrca}).}
\label{fig:theorcoolA}
\end{figure}
\begin{figure}
\includegraphics[width=0.51\textwidth]{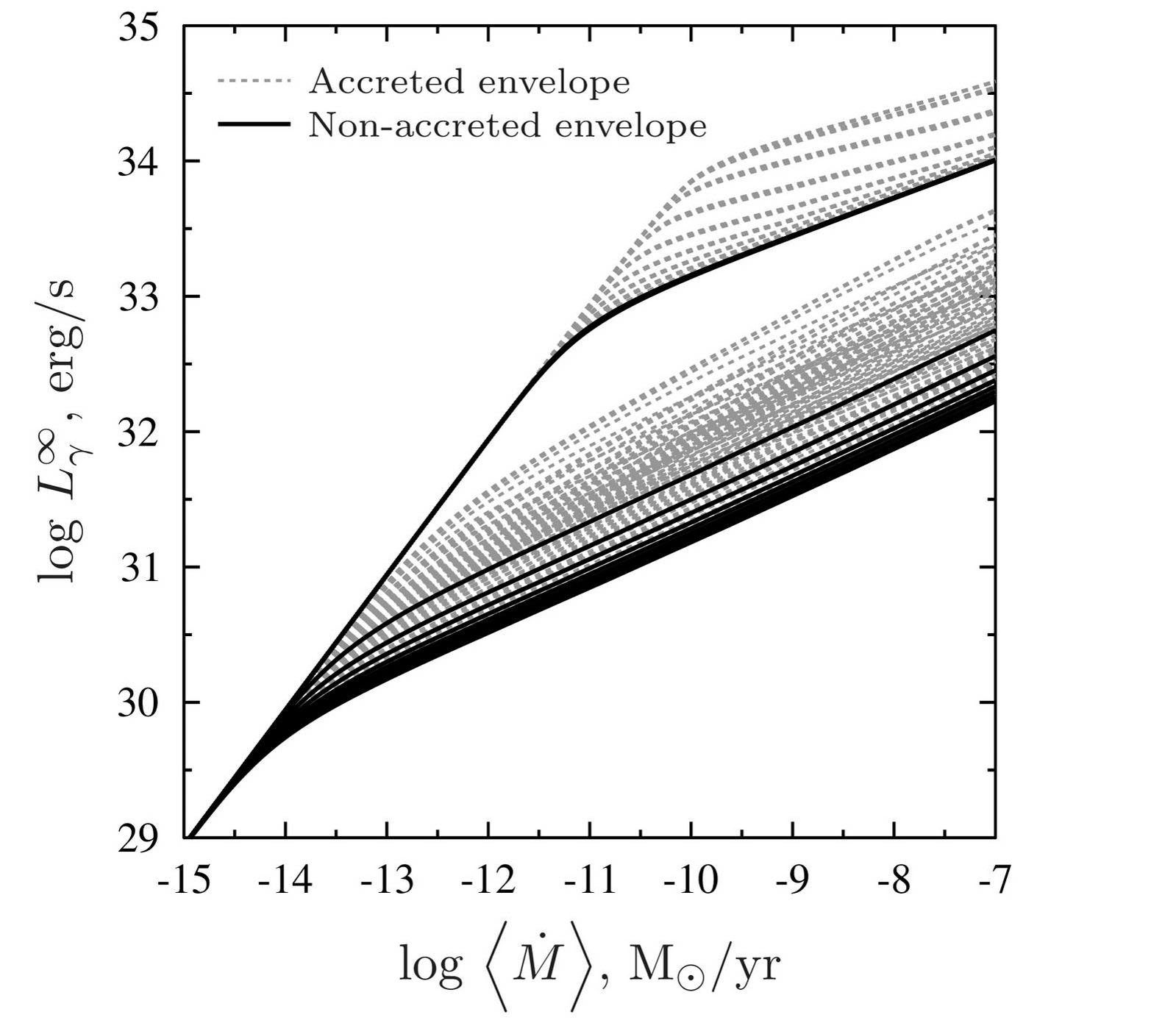}
\caption{A sequence of heating curves $L_\gamma
^\infty\left(\langle \dot{M} \rangle\right)$ of transiently accreting neutron
stars of masses $M=1.6\,\mathrm{M}_\odot-1.8\,\mathrm{M}_\odot$
($\Delta M=0.01\,{\mathrm{M}_\odot}$).
Solid curves correspond to iron heat blanket, dashed curves
are for heat blankets containing light elements of different
mass $\Delta M_{\rm le}
=10^{-k}\,M$, where $k$=7,8, \ldots , 16. The direct Urca threshold
is not broadened (see text for details).}
\label{fig:theorheatA}
\end{figure}

For example, Figs.\ \ref{fig:theorcoolNA} and \ref{fig:theorheatNA} show
sequences of cooling and heating curves of neutron stars of masses
from $M=1.1\,\mathrm{M}_\odot$ to $2.1\,\mathrm{M}_\odot$ (with the mass
step $\Delta M=0.01\,\mathrm{M}_\odot$); for
simplicity, the heat blanketing envelopes are assumed to be made of
iron. Indeed, this theory can in principle explain any cooling or
heating curve in the space between the upper ($1.1\,\mathrm{M}_\odot$) and
lower ($2.1\,\mathrm{M}_\odot$) curves, but the explanation might be unlikely.
For instance, all cooling curves of stars with $M\leq M_{\rm D}$ in
Fig.\ \ref{fig:theorcoolNA} merge actually into single (basic) cooling
curve which describes cooling of non-superfluid neutron stars via
the modified Urca process of neutrino emission (e.g., \citealt{YP04},
and references therein). However, at $M>M_{\rm D}$ the
power direct Urca process appears in an inner kernel of the star,
the star cools much faster, and becomes much colder than the stars
with $M\leq M_{\rm D}$. Therefore, we have actually two types of
cooling stars -- slowly cooling ($M\leq M_{\rm D}$, ``warm'') and rapidly
cooling ($M > M_{\rm D}$, ``cold'') ones separated by a ``gap'';
intermediate coolers are available but rather improbable.
Equivalently, we have two types of heating neutron stars (Fig.\ \ref{fig:theorheatNA}) -- sufficiently warm ($M\leq M_{\rm D}$) and much colder ($M > M_{\rm D}$) ones; intermediate stars are again rather improbable (the latter
circumstance is evident but is not widely known in the literature). The presence
of light elements in the heat blanketing envelope (i.e. accreted envelopes instead of pure iron)
somewhat reduce the ``gap''
and smooths the transition between ``cold'' and ``warm'' stars.
But as can be seen from Figs.\, \ref{fig:theorcoolA} and \ref{fig:theorheatA},
the presence of accreted matter cannot
actually merge two populations and fill in the ``gap''.
The existence of these two representative types of
cooling and heating neutron stars separated by a small amount of
intermediate sources formally contradicts the observations (Sect.\,
\ref{s:observa}, Figs.\ \ref{fig:cool} and \ref{fig:heat}). We will
show that it is actually not so.

Now we are ready to formulate statistical theory of the thermal
evolution of neutron stars. The stars in question are assumed to
have the same internal structure (EOS, neutrino emission properties)
but they can naturally have different parameters such as mass, the
amount of light elements in heat-blanketing envelopes, magnetic
fields, rotation, etc. In this situation instead of deterministic
cooling/heating curves in appropriate diagrams we can introduce
probabilistic (statistical) description, and discuss the probability
distributions to find a star in different places of a diagram.
These distributions can be obtained by averaging the cooling/heating
curves over statistical distributions of probabilistic parameters
such as masses $M$ and  the amount of light
elements in heat-blanketing envelopes. After this averaging, the
cooling/heating, that initially followed specific trajectories, is
replaced by statistical probabilities to find neutron stars at
different stages of their evolution.

In order to illustrate this scheme we simplify our consideration.
First, we neglect the effects of magnetic fields and rotation on
thermal states of cooling neutron stars and transiently accreting
neutron stars in XRTs. This seems to be a reasonably valid first
approximation. To study isolated cooling neutron stars and
transiently accreting neutron stars we introduce the distribution
functions over neutron star masses for these sources, $f_{\rm i}(M)$
and $f_{\rm a}(M)$. These functions are naturally different; the
masses of accreting neutron stars should be overall higher than
those of isolated neutron stars.

\begin{figure}
\includegraphics[width=0.51\textwidth]{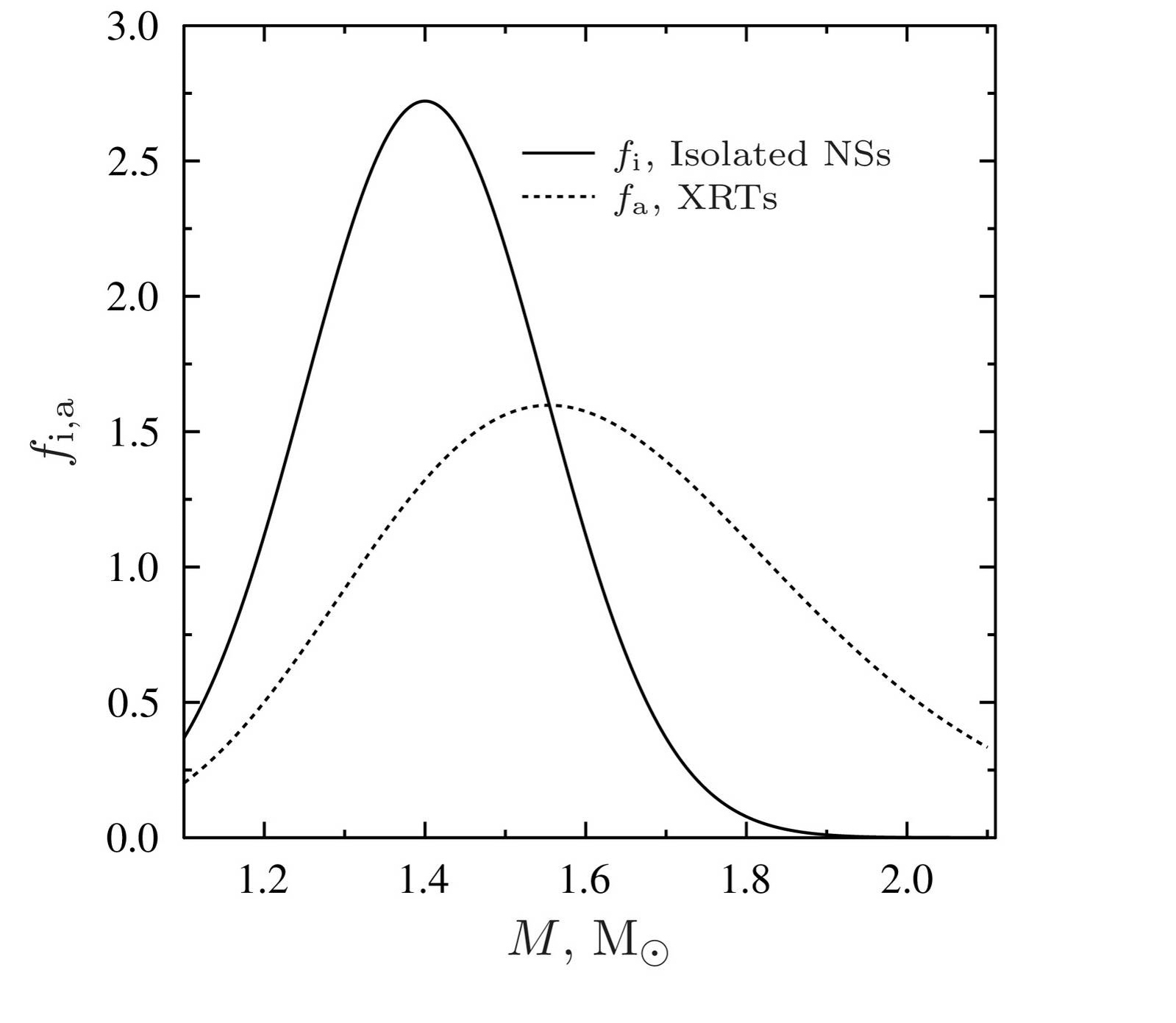}
\caption{Mass distributions of isolated neutron stars (solid curve)
and neutron stars in XRTs (dashed curve).
The parameters of the distributions are
$\mu_\mathrm{i}=1.4$~M$_\odot$, $\sigma_\mathrm{i}=0.15$~M$_\odot$;
$\mu_\mathrm{a}=0.47$  and
$\sigma_\mathrm{a}=0.17$ (see text for details). }
\label{fig:MDistrib}
\end{figure}

These distribution functions are taken in the form (Fig.\ \ref{fig:MDistrib})
\begin{equation}
\begin{split}
     &f_\mathrm{i}(M)=\frac{1}{N_\mathrm{i}}\frac{1}{\sqrt{2\pi}\,
        \sigma_\mathrm{i}}\exp{\left(-\frac{\left(M-\mu_\mathrm{i}\right)^2}
        {2\sigma_\mathrm{i}^2}\right)}, \\
    &f_\mathrm{a}(M)=\frac{1}{N_\mathrm{a}}\frac{1}
    {\sqrt{2\pi}\,M\sigma_\mathrm{a}}\exp{\left(-\frac{\left(\ln{\left[M/{\rm M}_\odot\right]}
    -\mu_\mathrm{a}\right)^2}{2\sigma_\mathrm{a}^2}\right)},
\end{split}
\label{e:mass-distrib}
\end{equation}
where $\sigma_\mathrm{i,a}$ and $\mu_\mathrm{i,a}$ are the
parameters of the distributions; $N_\mathrm{i,a}$ are normalization
factors, which rescale these distribution to the finite mass range
from $1.1\,\mathrm{M}_\odot$ to $2.1\,\mathrm{M}_\odot$. For
$M<1.1\,\mathrm{M}_\odot$ and  $M>2.1\,\mathrm{M}_\odot$ these
distribution functions are artificially set to zero. Note that for
the normal distribution function $f_\mathrm{i}(M)$, the
parameter $\mu_\mathrm{i}$ is the most probable mass. However, for
the lognormal distribution $f_\mathrm{a}(M)$ the most probable mass
is equal to
M$_\odot\exp{\left(\mu_\mathrm{a}-\sigma_\mathrm{a}^2\right)}$.
After some test runs we have taken the distributions with
$\mu_\mathrm{i}=1.4$~M$_\odot$, $\sigma_\mathrm{i}=0.15$~M$_\odot$;
$\mu_\mathrm{a}=0.47$  and $\sigma_\mathrm{a}=0.17$; the most
probable mass for the accreting neutron stars is 1.55\,M$_\odot$. It
seems these functions do not contradict the data and theoretical
expectations (e.g., \citealt{Kiziltan_etal13}) but they are
definitely not unique. It is important that accreting neutron
stars are overall heavier as a natural result of accretion.

The heat transparency of the blanketing envelope is determined by
the mass $\Delta M_{\rm le}$ of light elements (mainly, hydrogen and
helium) in these envelopes. The higher $\Delta M_{\rm le}$, the
larger thermal conductivity in the envelope, and the higher $T_{\rm
s}$ for a given internal temperature of the star (e.g.,
\citealt{Potekhin_etal97}). However, $\Delta M_{\rm le}$ cannot be
larger than $\Delta M_{\rm le\,max}\approx 10^{-7}\,M$ because at
formally larger $\Delta M_{\rm le}$ the light elements at the bottom
of the heat blanketing envelope transform into heavier ones due to
beta captures and pycnonuclear reactions. We will consider $\Delta
M_{\rm le} \leq \Delta M_{\rm le\,max}$ as a random quantity which
is characterized by a distribution function $f_{\rm acc}(\Delta
M_{\rm le})$. By way of illustration, in calculations we take
\begin{equation}
   f_{\rm acc}(\Delta M_{\rm le})={\rm const}~~~{\rm at~~~}
    \Delta M_{\rm le}\leq 10^{-7}M.
\label{e:f-DeltaM}
\end{equation}
The $f_{\rm acc}(\Delta M_{\rm le})$ distribution is highly
uncertain; we take (\ref{e:f-DeltaM}) to show the range of effects
such distributions can produce.

\begin{figure}
\includegraphics[width=0.51\textwidth]{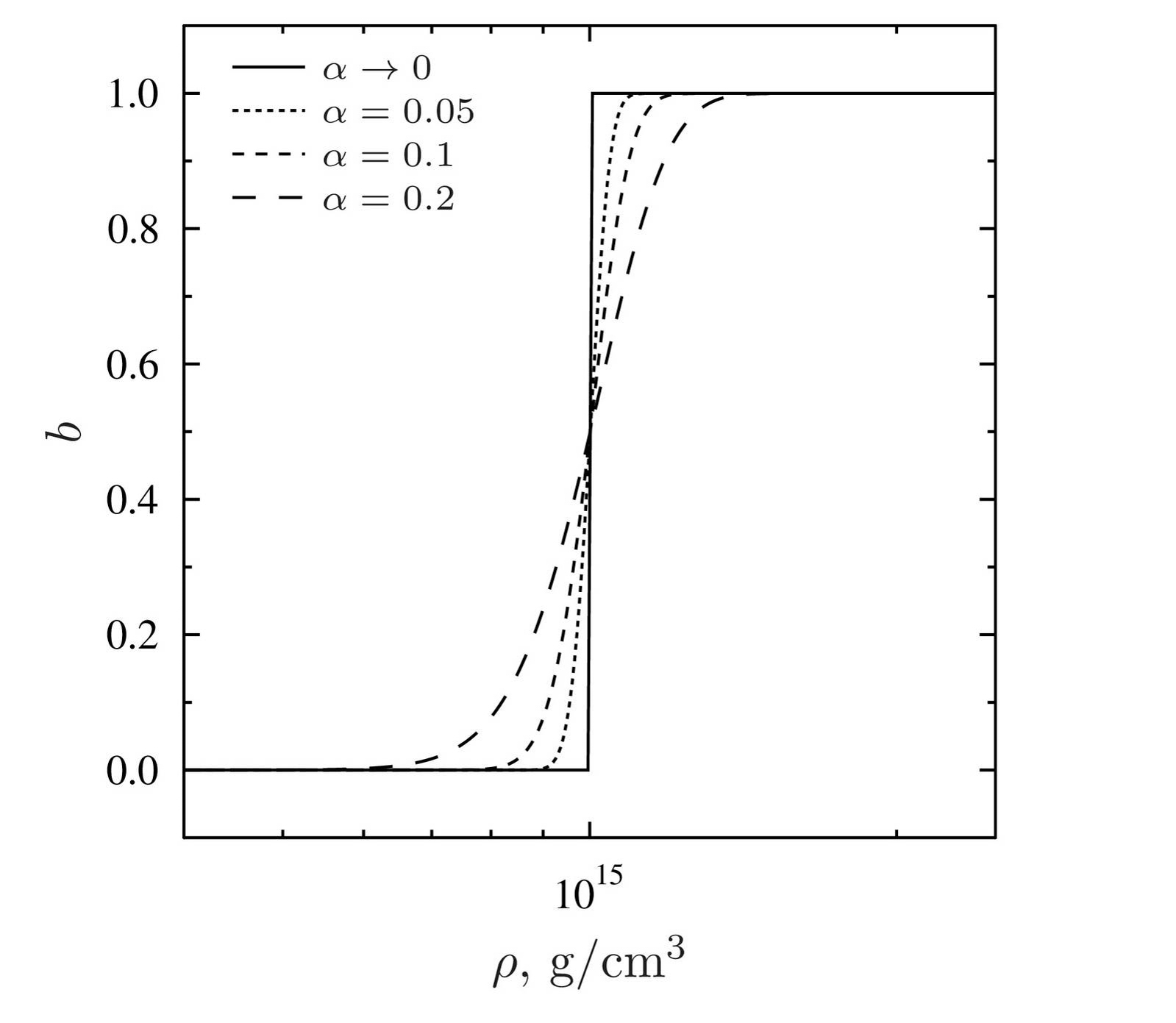}
\caption{Function $b$ versus  $\rho$, Eq.\ \eqref{e:broadD},
which approximates broadening of the electronic direct Urca
threshold $\rho_{\rm D}$ for three values of $\alpha=$ 0.05, 0.1 and
0.2. The solid line ($\alpha \to 0$) corresponds to no broadening at
all (see text for details).} \label{fig:BroadFunc}
\end{figure}

Our cooling code \citep{GYP01} allows us to take into account the
effects of superfluidity on thermal evolution of neutron stars. To
reduce the number of variable parameters, we employ a
semiphenomenological approach (although we mention some effects of
superfluidity in Section \ref{s:broadUrca}). In particular, we will
broad out artificially a step-like density dependence of the
neutrino emissivity $Q_{\rm D}$ provided by the direct Urca process
\citep{LPPH91}. In the absence of superfluidity the direct Urca
process switches on sharply with increasing density, from $Q_{\rm
D}=0$ at $\rho < \rho_{\rm D}$ to finite $Q_{\rm D}$ at $\rho \geq
\rho_{\rm D}$ (solid curve in Fig.\ \ref{fig:BroadFunc}). Moreover,
in our model HHJ EOS, superdense matter of neutron star cores
consists of neutrons with admixture of protons, electrons and muons,
and we have direct Urca processes of two types, electronic and
muonic ones (e.g., \citealt{YKGH01}). Accordingly, we have two
density thresholds for the onset of the electronic and muonic
processes (and the emissivity of both processes -- if open -- is the
same). The density threshold for the muonic process is always higher
than for the electronic one. Accordingly, when we increase $M$ (or
$\rho_{\rm c}$) the electronic direct Urca always switches on first,
sharply (by 6--7 orders of magnitude) increases the neutrino
luminosity of the star, and appears to be the leading one. The
switch-on of the muonic process with further increase of $M$ or
$\rho_{\rm c}$ is relatively unimportant (although included properly
in the calculations). It is well known (see below) that a sharp
step-like onset of the direct Urca process is incompatible with
observations. One needs to broaden the direct Urca threshold. We
will include this broadening on a phenomenological level by
multiplying the electronic and muonic neutrino emissivities by a
broadening factor $b$. For instance, for the electronic direct Urca
we take (Fig.\ \ref {fig:BroadFunc})
\begin{equation}
    Q_\mathrm{D}=Q_\mathrm{D0}\,b(x),
        \quad b(x)= 0.5\left[1+\erf{\left(x\right)}\right],
\label{e:broadD}
\end{equation}
where $Q_{\rm D0}$ is the threshold emissivity, $b=b(x)$,
$x=(\rho-\rho_{\rm D})/(\alpha \rho_{\rm D})$, erf($x$) is the
standard error function, so that $b(x) \to 0$ at $x \to - \infty$
and $b(x) \to 1$ at $x \to \infty$, and $\alpha$ is a parameter
assumed to be small, $\alpha \ll 1$ (see Fig.\
\ref{fig:BroadFunc}). This parameter determines a narrow range of
densities $|\rho-\rho_{\rm D}|\sim \alpha \rho_{\rm D}$ in which the
direct Urca process gains its strength. Similar broadening is
introduced for the muonic direct Urca process but it does not affect
significantly our results.

For example, the broadening of the direct Urca threshold can be
provided by proton superfluidity (e.g., \citealt{YKGH01}). This
superfluidity (due to singlet-state pairing of protons) is
characterized by the proton critical temperature $T_{\rm cp}(\rho)$
(e.g., \citealt{LS01}). The critical temperatures are very model
dependent, with a large scatter of theoretical $T_{\rm cp}(\rho)$,
so that it is instructive to not to rely on specific theoretical
models but to consider $T_{\rm cp}(\rho)$ on phenomenological level.
One can expect that proton superfluidity is strong in the outer core
of the neutron star (with $T_{\rm cp}(\rho)\gtrsim 3 \times 10^9$~K)
but becomes weaker or disappears entirely in the inner core, at a
few nuclear matter densities. As long as it is strong, it greatly
suppresses the direct Urca process (even if the direct Urca is
formally allowed) by the presence of a large gap in the energy
spectrum of protons. When proton superfluidity becomes weaker with
growing $\rho$, the superfluid suppression is removed and the direct
Urca becomes very powerful. It switches on after exceeding some
threshold density, but not very sharply, as if the threshold is
broadened.

In addition to the nucleonic direct Urca process, there could be
weaker processes of fast neutrino emission produced, for instance,
by the presence of pion or kaon condensates in inner cores of
neutron stars (e.g., \citealt{YKGH01}, and references therein).
These processes are known to be important if the direct Urca process
itself is forbidden or greatly suppressed. We will consider such
situations in an approximate manner by multiplying the emissivity
due to the direct Urca process by a factor $\beta$, where $\beta
\sim 10^{-2}$ or $10^{-4}$ imitate the presence of pion or
kaon condensations, respectively.

All elements of cooling/heating theory of neutron stars
 employed in our calculations are not new. The new element
consists in implementing statistical theory (distributions over the neutron star masses and over the amount of light elements in the
heat blanketing envelopes).

\section{Results and discussion}
\label{s:res}

\subsection{Broadening direct Urca threshold}
\label{s:broadUrca}

\begin{figure}
\includegraphics[width=0.51\textwidth]{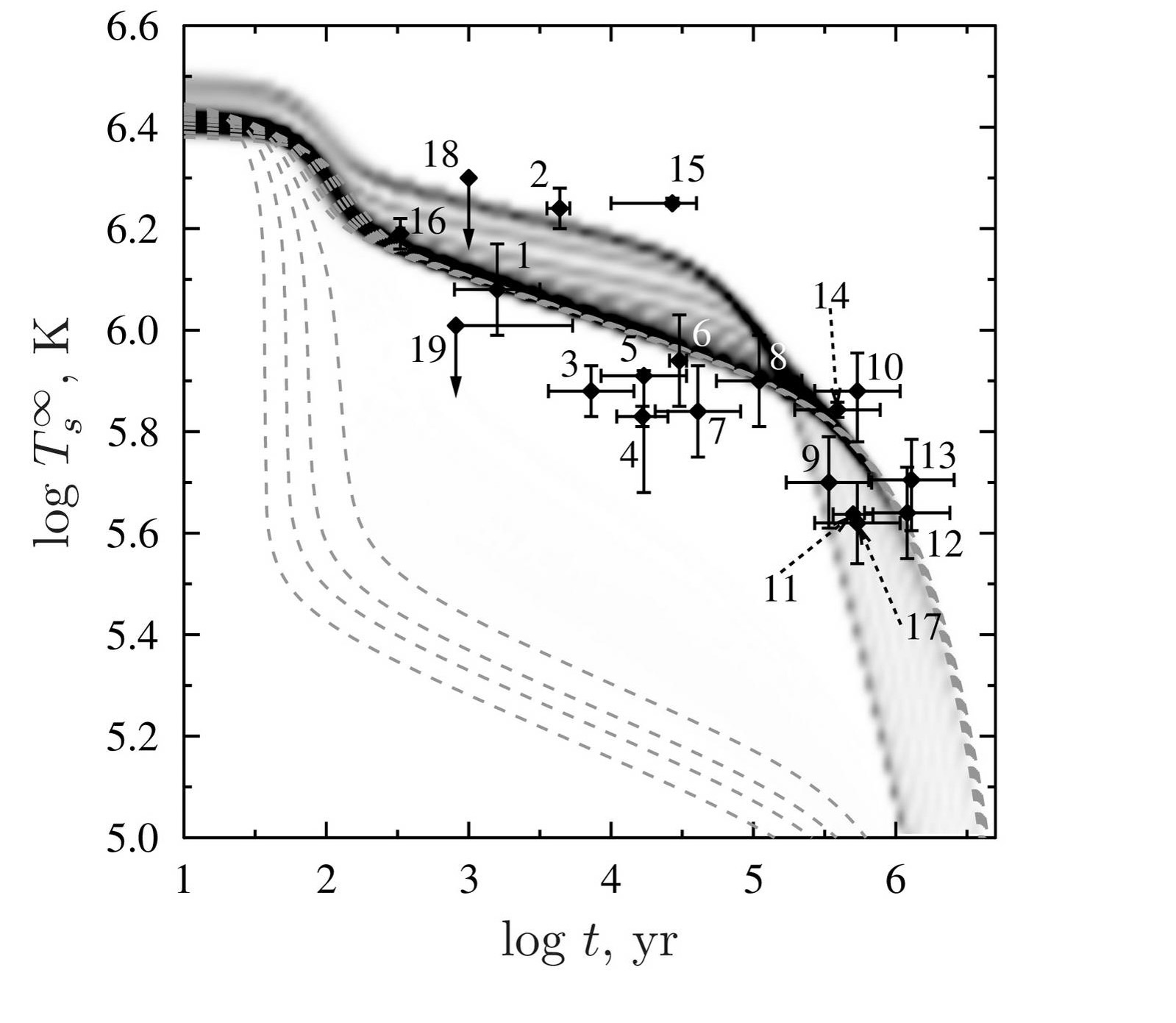}
\caption{Probability to find a cooling isolated neutron star in
different places of the $T_{\rm s}^\infty-t$ plane compared with
observations (Fig.\ \ref{fig:cool}). The distributions over neutron
star masses and over the amount of light elements in surface layers
are given by equations (\ref{e:mass-distrib}) and
(\ref{e:f-DeltaM}), respectively. Dashed lines show 11 ``reference''
cooling curves for stars with iron envelopes and masses
$M=1.1,\,1.2\,, \ldots, 2.1\,\mathrm{M}_\odot$. The direct Urca
process threshold is not broadened. See text for details.}
\label{fig:coolD_1000}
\end{figure}

\begin{figure}
\includegraphics[width=0.51\textwidth]{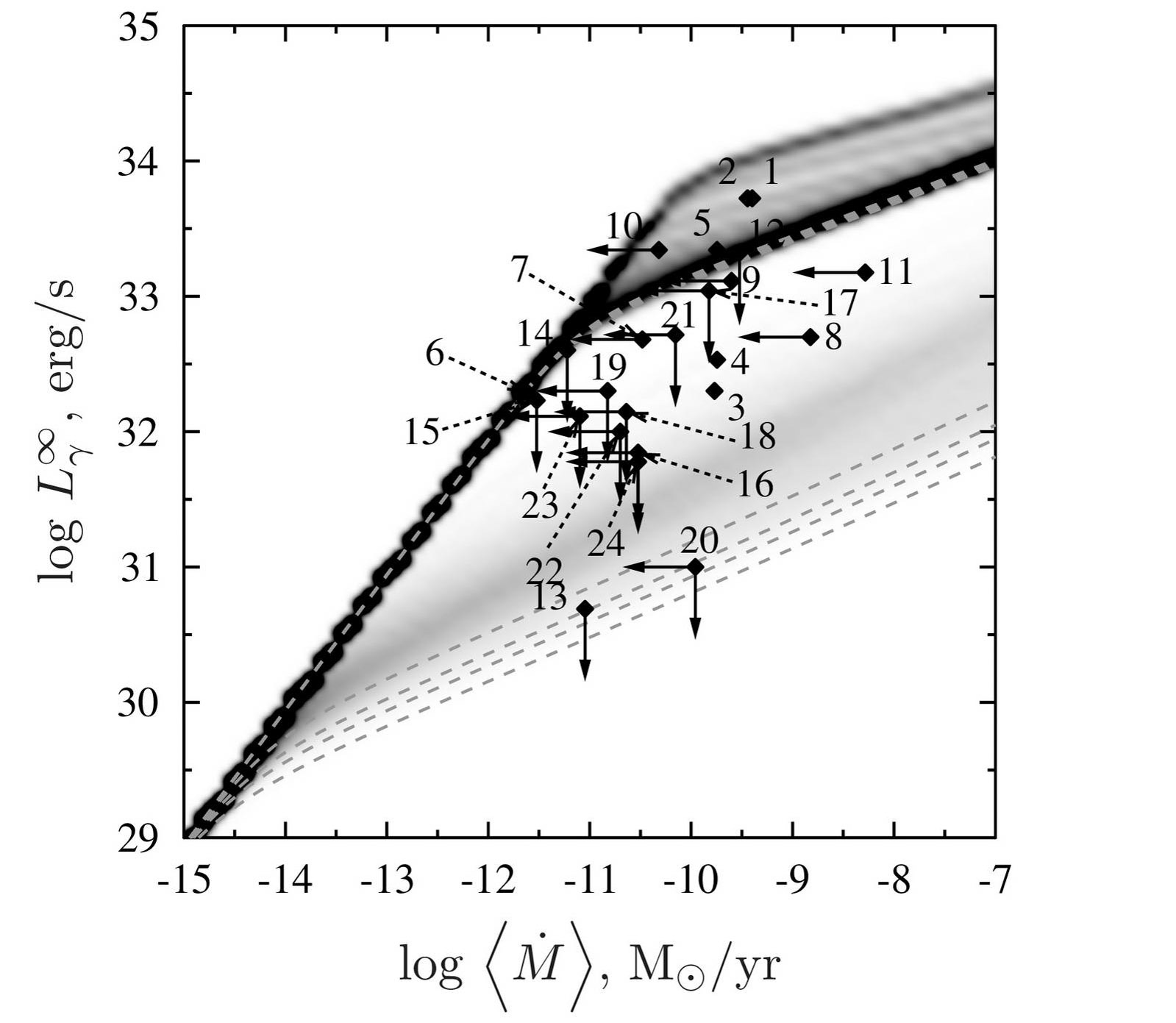}
\caption{Probability to find a transiently accreting
neutron star in different places of the $L_{\gamma}^\infty-\langle
\dot{M} \rangle$ plane compared with observations (Fig.\
\ref{fig:heat}). 
Dashed lines show 11 ``reference'' heating curves for stars with
iron envelopes and masses $M=1.1\,, 1.2\,, \ldots ,
2.1\,\mathrm{M}_\odot$. The direct Urca threshold is not broadened.}
\label{fig:heatD_1000}
\end{figure}

Figs.\ \ref{fig:coolD_1000} and \ref{fig:heatD_1000} show
calculated probabilities to find isolated cooling neutron stars and
transiently accreting neutron stars in different places of the
$T_{\rm s}^\infty-t$ and $L_{\gamma}^\infty-\langle \dot{M} \rangle$
diagrams, respectively. The results are compared with observations
(Figs. \ref{fig:cool} and \ref{fig:heat}). The probabilities are
calculated by averaging over neutron star masses in accordance with
(\ref{e:mass-distrib}) and over the amount of light elements in the
heat blanketing envelopes, equation (\ref{e:f-DeltaM}).
The probability distribution is presented by grayscaling (in relative units).
The denser the scaling, the large the probability. White regions
refer to zero or very low probability.

\begin{figure}
\includegraphics[width=0.51\textwidth]{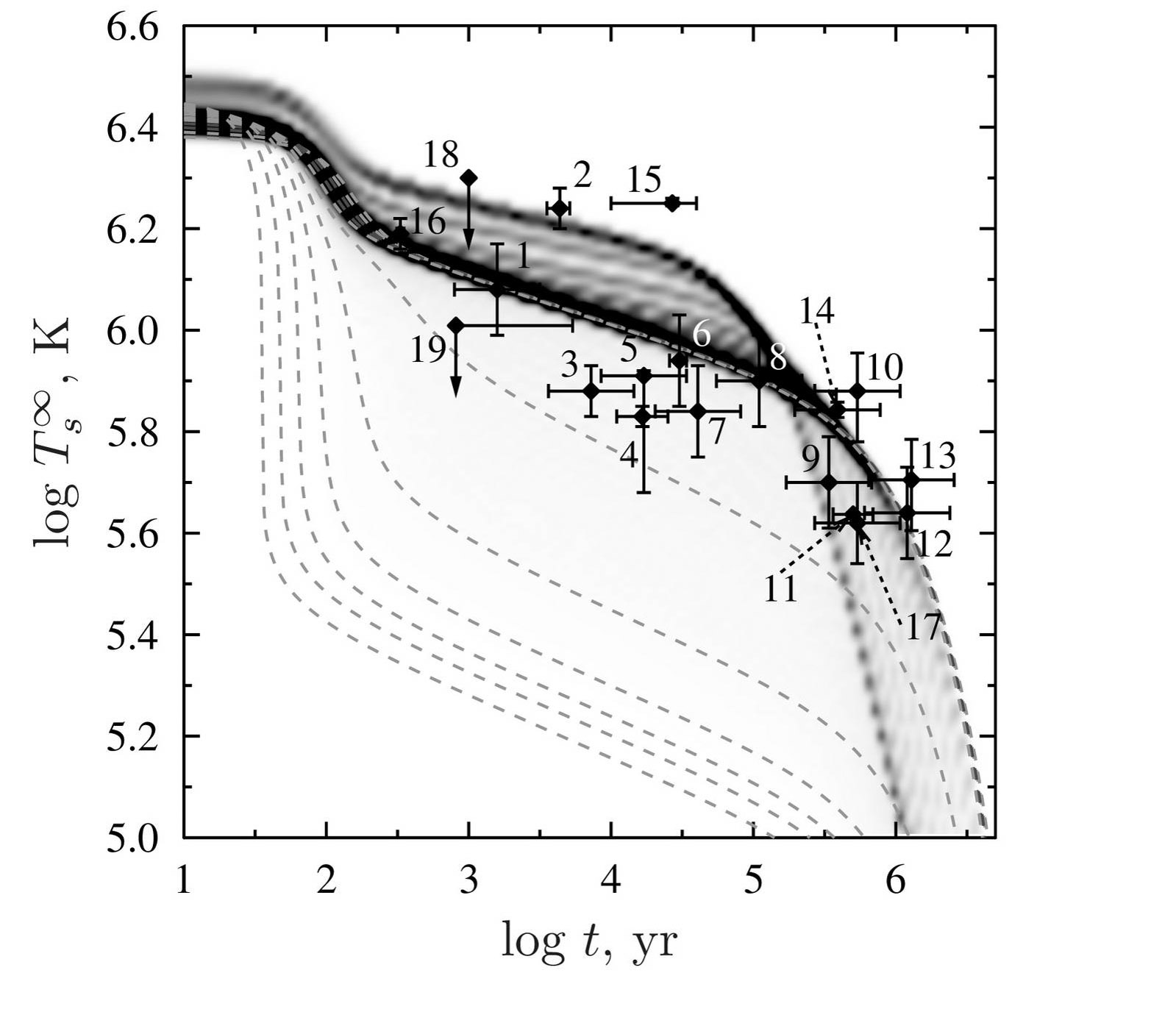}
\caption{Same as in Fig.\ \ref{fig:coolD_1000}
but with the direct Urca threshold broadened,
according to equation (\ref{e:broadD}) with $\alpha=0.05$.}
\label{fig:coolD_20}
\end{figure}

\begin{figure}
\includegraphics[width=0.51\textwidth]{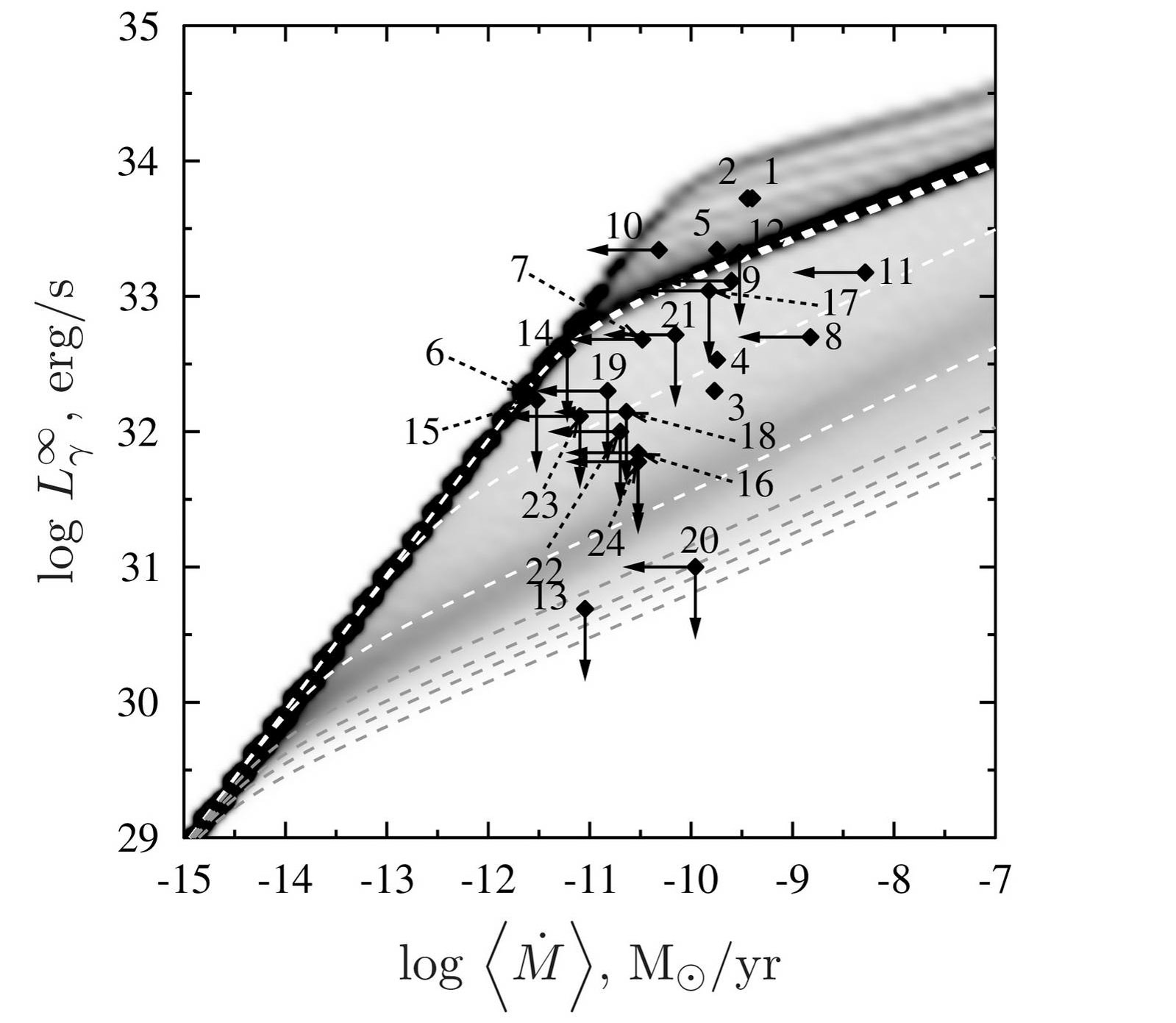}
\caption{Same as in Fig.\ \ref{fig:heatD_1000}
but with slightly broadened direct Urca threshold, $\alpha=0.05$.}
\label{fig:heatD_20}
\end{figure}

In Figs.\ \ref{fig:coolD_1000} and \ref{fig:heatD_1000} the
threshold of the direct Urca process is not broadened (the
solid line in Fig.\ \ref{fig:BroadFunc}). Because of the sharp contrast
of neutrino luminosities of neutron stars with open and closed
direct Urca process, the averaging (\ref{e:f-DeltaM}) does not
greatly affect the probabilities to find neutron stars in different
places of respective diagrams. This averaging slightly broadens the
distributions of rather warm neutron stars ($M \leq M_{\rm D}$) and
rather cold ones ($M > M_{\rm D}$) but does not remove large ``gap''
between them. It evidently contradicts the observations of cooling
and heating neutron stars.

\begin{figure}
\includegraphics[width=0.51\textwidth]{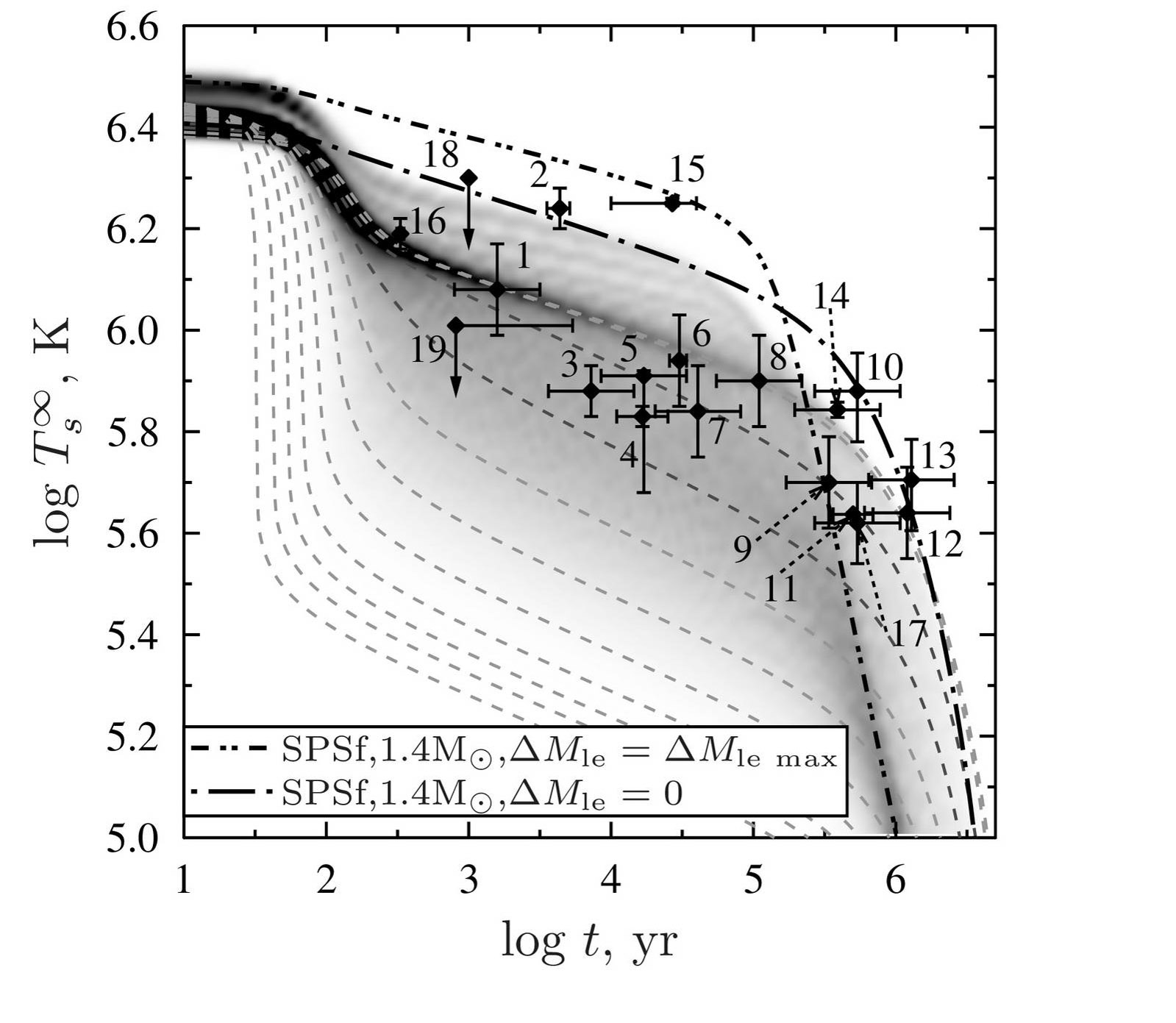}
\caption{Same as in Fig.\ \ref{fig:coolD_1000}
but with the direct Urca threshold broadened in the way ($\alpha=0.1$) to
achieve agreement with the observational data. The additional dot-dashed line is for
the 1.4\,M$_\odot$ star with strong proton superfluidity in the core and
iron envelope; the dashed-double-dot line is for the same star but with
the maximum amount of light elements in the heat blanket (see text for details).}
\label{fig:coolD_10}
\end{figure}

\begin{figure}
\includegraphics[width=0.51\textwidth]{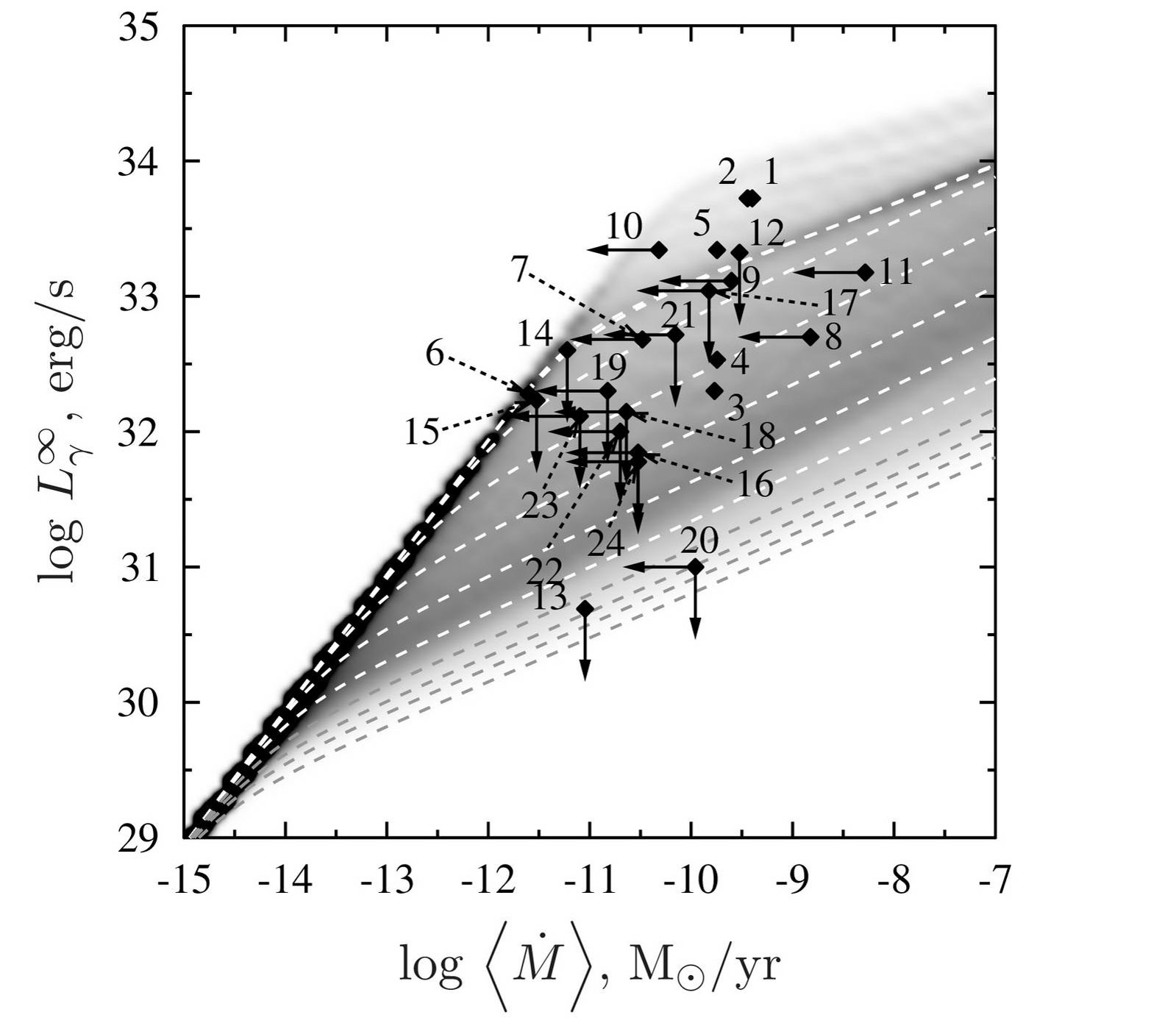}
\caption{Same as in Fig.\ \ref{fig:heatD_1000}
but with the direct Urca threshold broadened  in the way ($\alpha=0.1$) to
agree with the observational data.}
\label{fig:heatD_10}
\end{figure}

As the next stage let us slightly broaden the direct Urca threshold
taking $\alpha=0.05$ in equation (\ref{e:broadD}) (the
dotted line in Fig.\ \ref{fig:BroadFunc}). The results are plotted
in Figs.\ \ref{fig:coolD_20} and \ref{fig:heatD_20}. As seen from
Fig.\ \ref{fig:coolD_20}, such a broadening is insufficient to merge
the ``warm'' and ``cold'' populations of cooling neutron stars
(although on these grayscale images it is difficult to see the
difference in probability diftributions in  Figs.\ \ref{fig:coolD_1000} and \ref{fig:coolD_20}, the
difference in ``reference'' curves is clearly seen). The first
glance at Fig. \ref{fig:heatD_20} may give an impression that such a
small broadening is sufficient for transiently accreting neutron
stars in XRTs but it is not so. A thorough examination reveals that the
probability density is too high in the region of a few ``warmest''
sources (1 and 2); in addition, it is too low in the ``dense''
region of ``intermediate'' sources such as 19, 21 and 23. One can
also notice the non-uniformity of ``reference'' curves (especially
if compared to proper threshold broadening; see below). These facts indicate that $\alpha=0.05$
provides insufficient broadening of the direct Urca threshold.

Now we broaden the direct Urca threshold in a such way that
probability density coincides with observational data for isolated
and accreting neutron stars. To achieve this we take $\alpha=0.1$ in
(\ref{e:broadD}); see the short-dashed line in Fig.\
\ref{fig:BroadFunc}. The results are plotted in Figs.\
\ref{fig:coolD_10} and \ref{fig:heatD_10} and seem to be in good agreement
with observations; the ``gap'' is completely removed; the ``warm''
and ``cold'' neutron star populations merge into one population as
the data prescribe.

In addition, in Fig. \ref{fig:coolD_10}
we plot two cooling curves for the 1.4\,M$_\odot$ star.
The dash-and-dot curve is for the case when the star has the iron heat
blanket and strong proton superfluidity inside (with the critical
temperature $T_{\rm cp}(\rho) \gtrsim 3 \times 10^9$~K over the core).
This superfluifity suppresses the modified Urca process
(e.g., \citealt{YP04}) and makes the star warmer. For stars
of age $t \lesssim 10^5$ yr, it produces nearly the same affect
on the cooling as the heat blanketing envelope made of
light elements. The dashed-double-dot curve is for the same proton superfluidity
but for the heat blanket with the maximum amount of light elements.
The star becomes even warmer and demonstrates exceptionally slow cooling
which is consistent even with observations of XMMU J1732--3445 (source 15), the hottest
isolated neutron star (for its age). These two curves are presented for
illustration, to demonstrate that the cooling theory is able to explain
all the sources. These curves have not been included in the calculations
of the probability distribution.

\begin{figure}
\includegraphics[width=0.51\textwidth]{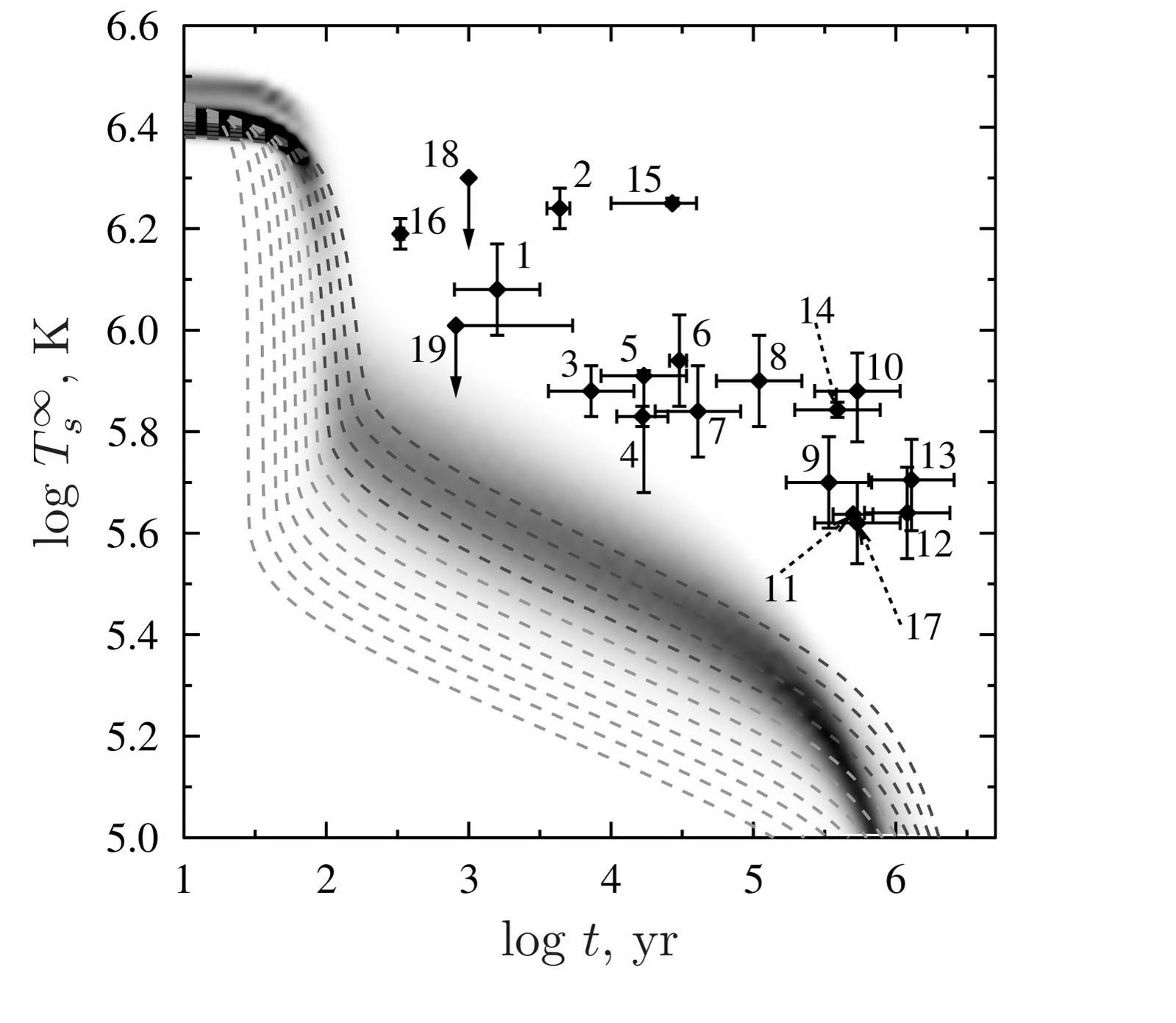}
\caption{Same as in Figs.\ \ref{fig:coolD_1000}
and \ref{fig:coolD_10} but with the direct Urca threshold broadened
too much, with $\alpha=0.2$.}
\label{fig:coolD_5}
\end{figure}

\begin{figure}
\includegraphics[width=0.51\textwidth]{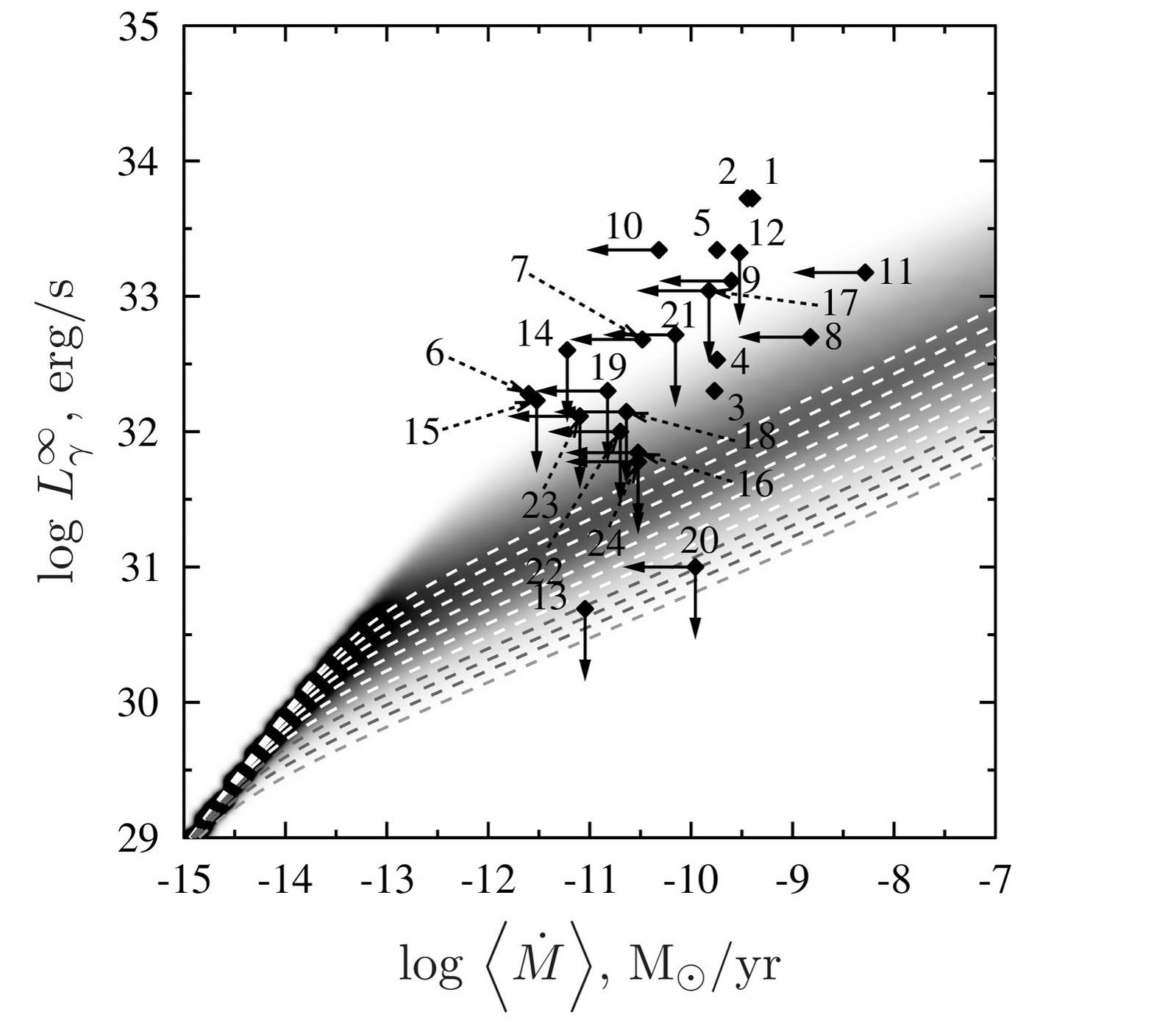}
\caption{Same as in Figs.\ \ref{fig:heatD_1000} and
\ref{fig:heatD_10} but with the overbroadened direct Urca
threshold
($\alpha=0.2$).}%
\label{fig:heatD_5}
\end{figure}

Finally, let us broaden the direct Urca threshold even more, taking
$\alpha=0.2$ in equation (\ref{e:broadD}); see the long-dashed line
in Fig.\ \ref{fig:BroadFunc}. These results are plotted in Figs.\
\ref{fig:coolD_5} and \ref{fig:heatD_5}. All cooling/heating curves
shift towards the ``cooler'' part of the cooling/heating plane
because now the direct Urca process operates even in low mass stars.
This situation evidently contradicts the observations.

It has been a longstanding problem to interpret the
observations of the transiently accreting source SAX 1808.4--3654
(source 13). It seems to contain a very cold star whose observations
in quiescent periods require the operation of direct Urca process,
while the observed isolated middle-aged neutron stars do not
contain such a cold source. A natural explanation of this phenomenon
within the present model is that the neutron star in SAX
1808.4--3654 is sufficiently massive; its mass $M$ is larger than
typical masses available in the mass distribution of isolated
neutron stars. Tuning the parameters of the mass distributions
$f_{\rm i}(M)$ and $f_{\rm a}(M)$ we naturally explain the effect
(Figs.\ \ref{fig:coolD_10} and \ref{fig:heatD_10}). In this model
the very cold isolated middle-aged neutron stars can (in principle)
exist in nature (in the tail of the mass distribution $f_{\rm
i}(M)$) but as a very rare phenomenon.

Therefore, the model presented in Figs.\ \ref{fig:coolD_10}
and \ref{fig:heatD_10} seems reasonable to explain the
available observations of cooling isolated and transiently accreting
neutron stars. The model requires a moderate broadening of the
direct Urca threshold and realistic masses of isolated and accreting
neutron stars. The broadening can be provided, for instance, by
proton superfluidity in the neutron star core as discussed above.

\subsection{Less enhanced neutrino cooling}
\label{s:pion-kaon}

Now consider the question if we can explain the data assuming that
the direct Urca process is forbidden in stars of all masses but less
powerful process of neutrino emission enhanced, for instance, by
pion or kaon condensation in inner cores of massive neutron stars is
present. To simulate such models we multiply the neutrino emissivity
$Q_{\rm D}$ due to the direct Urca process by a factor $\beta$,
where $\beta \sim 10^{-2}$ would be typical for pion condensation
and $\beta \sim 10^{-4}$ for kaon condensation.

\begin{figure}
\includegraphics[width=0.51\textwidth]{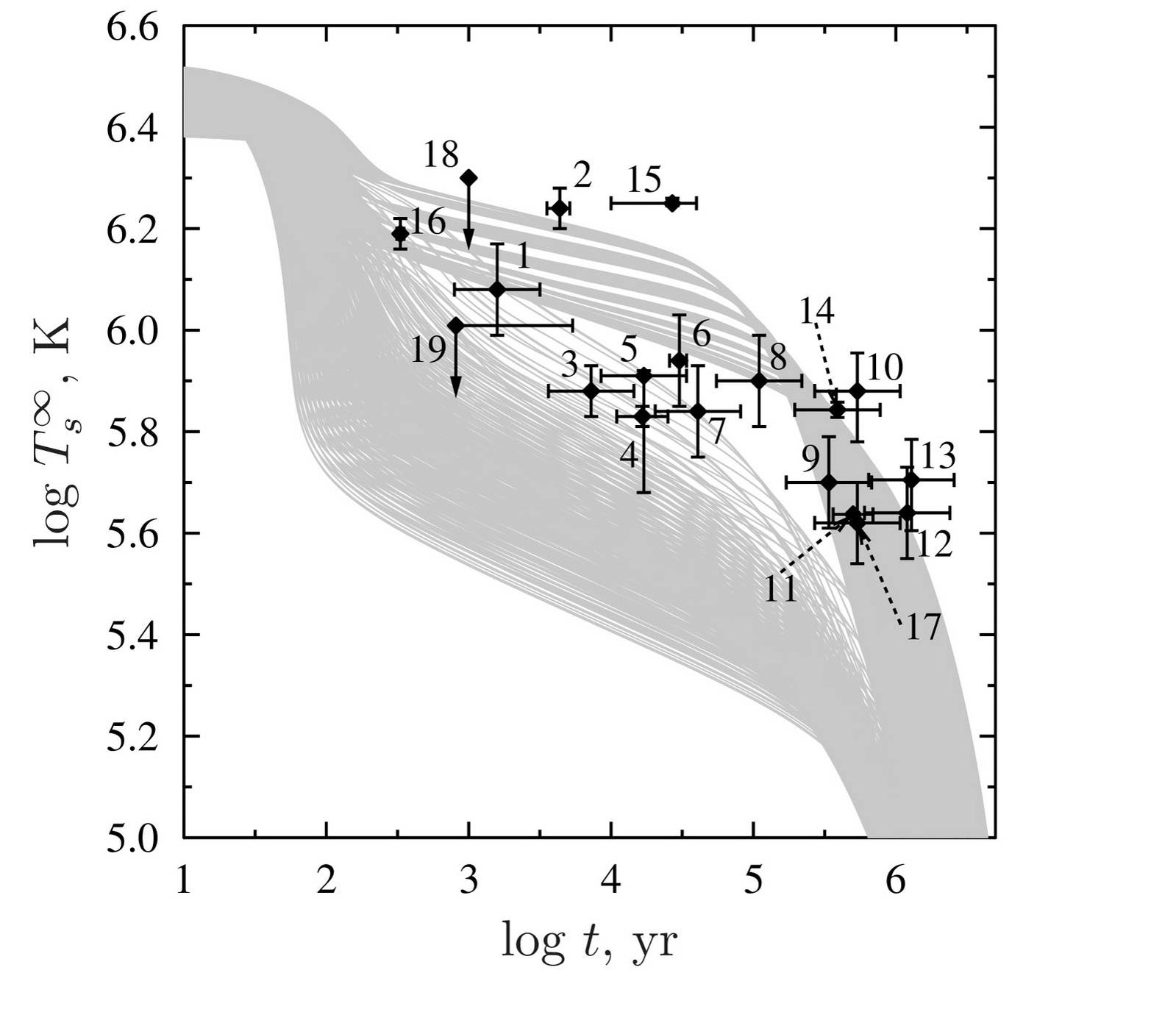}
\caption{A sequence of cooling curves $T_{\rm
s}^\infty(t)$ for neutron stars of masses
$M=1.11\,\mathrm{M}_\odot-2.09\,\mathrm{M}_\odot$ with mass step
$\Delta M=0.01\,{\mathrm{M}_\odot}$ and with different, iron and
accreted, heat blankets. The threshold for the
enhanced neutrino emission ($\beta=10^{-2}$) is not broadened  (see text for details).}
\label{fig:coolCR_100}
\end{figure}

\begin{figure}
\includegraphics[width=0.51\textwidth]{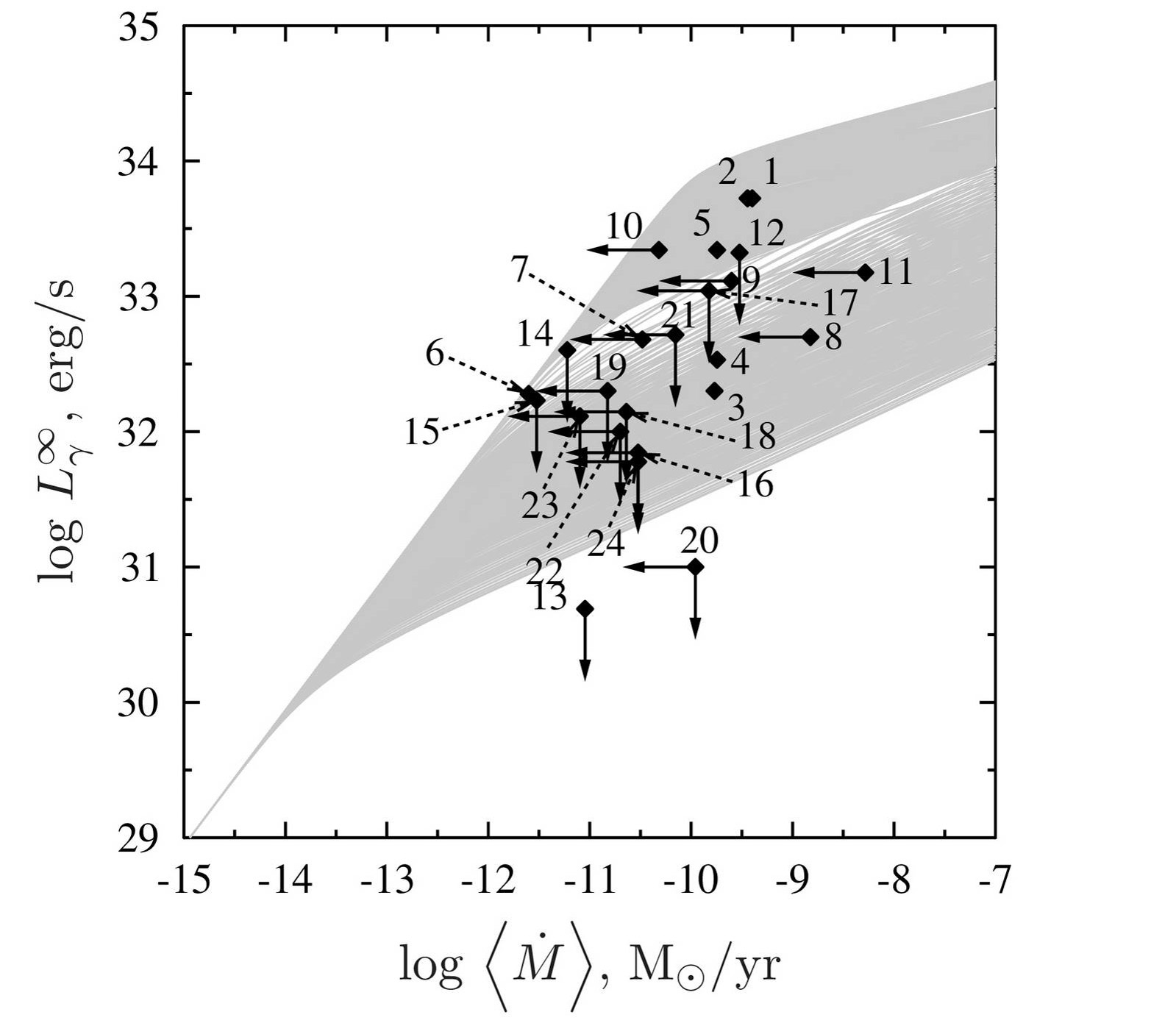}
\caption{A sequence of heating curves $L_\gamma
^\infty\left(\langle \dot{M} \rangle\right)$ of transiently accreting neutron
stars of masses $M=1.11\,\mathrm{M}_\odot-2.09\,\mathrm{M}_\odot$ with mass
step $\Delta M=0.01\,{\mathrm{M}_\odot}$ and with
different, iron and accreted, heat blankets. The threshold for the enhanced
neutrino emission ($\beta=10^{-2}$) is not broadened.
The coldest sources (13 and 20) contradict this model.
See text for details.}
\label{fig:heatCR_100}
\end{figure}

We start with the case of $\beta=10^{-2}$ (Figs.\ \ref{fig:coolCR_100} and \ref{fig:heatCR_100}). This case is
qualitatively similar to the case of open direct Urca process
(Section \ref{s:broadUrca}). Without broadening the threshold
$\rho_{\rm D}$ of the enhanced neutrino emission we would have two
distinct populations of rather warm ($M \leq M_{\rm D}$) and cold
($M>M_{\rm D}$) neutron stars separated by a wide ``gap'' (in
disagreement with the observations). However, the gap would be narrower
than in Section \ref{s:broadUrca} and colder stars would be warmer.
Averaging over the distribution of masses of light elements in
the heat blanketing envelope somewhat broadens both populations but
the effect is again rather insignificant. If we introduce some
broadening of the threshold for the enhanced emission, the two
populations of stars will merge into one population. However, it is
most important that now the transiently accreting massive neutron
stars would be warmer and we would never be able to explain the
existence of the ultracold neutron star in SAX 1808.4--3658 (see Fig.\  \ref{fig:heatCR_100}).
This star can be explained {\it only if} the direct Urca process
operates in a massive neutron star. Considering only the isolated
cooling neutron stars (and disregarding the transiently accreting
ones) we would be able to explain all the sources by setting
appropriate value for $\alpha$.

\begin{figure}
\includegraphics[width=0.51\textwidth]{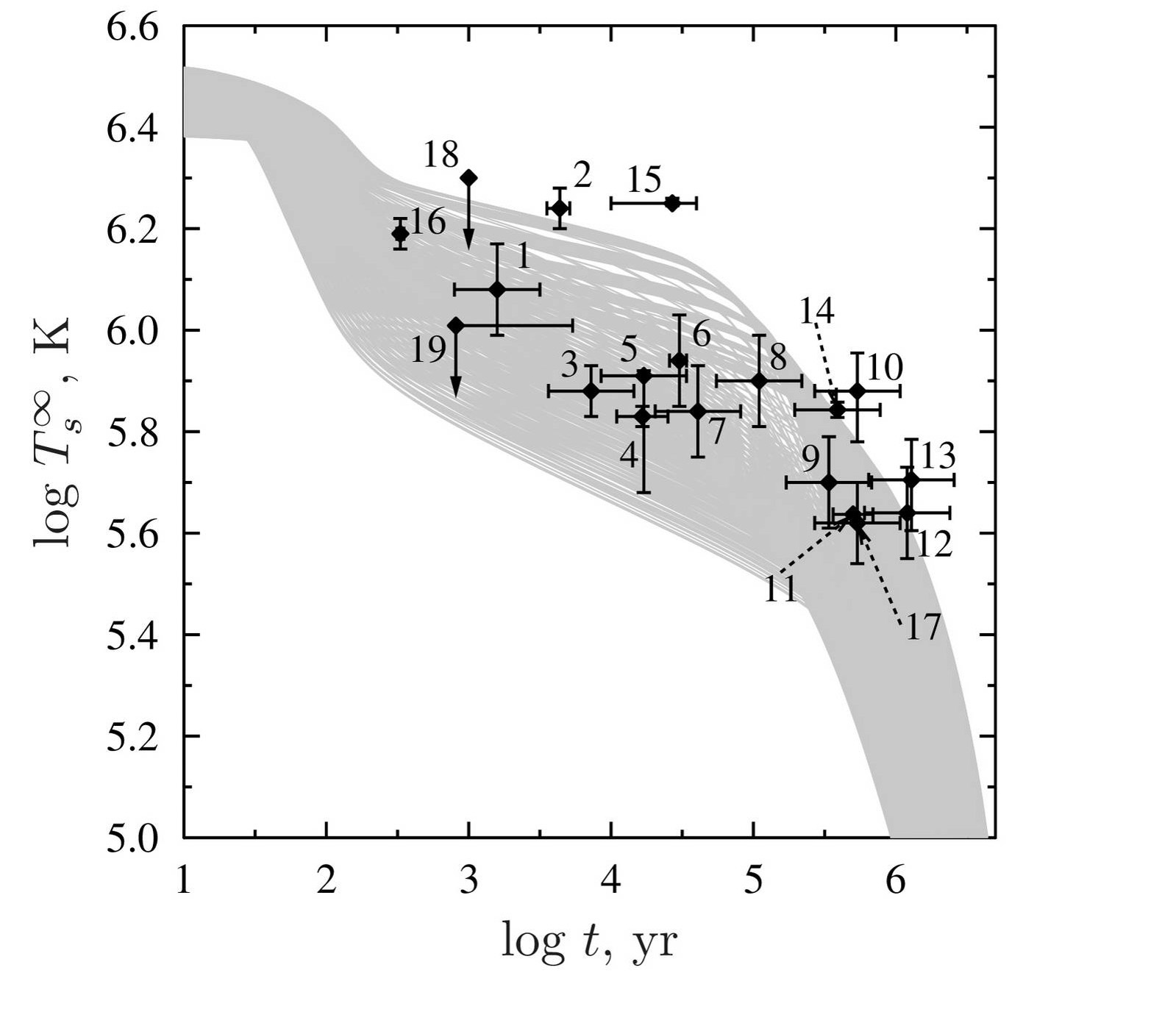}
\caption{Same as in Fig.\ \ref{fig:coolCR_100}, but the enhanced
neutrino emissivity is multiplied by a factor of  $\beta=10^{-4}$.  The ``gap''
between the two populsations of cooling stars almost disappears.}
\label{fig:coolCR_10000}
\end{figure}

\begin{figure}
\includegraphics[width=0.51\textwidth]{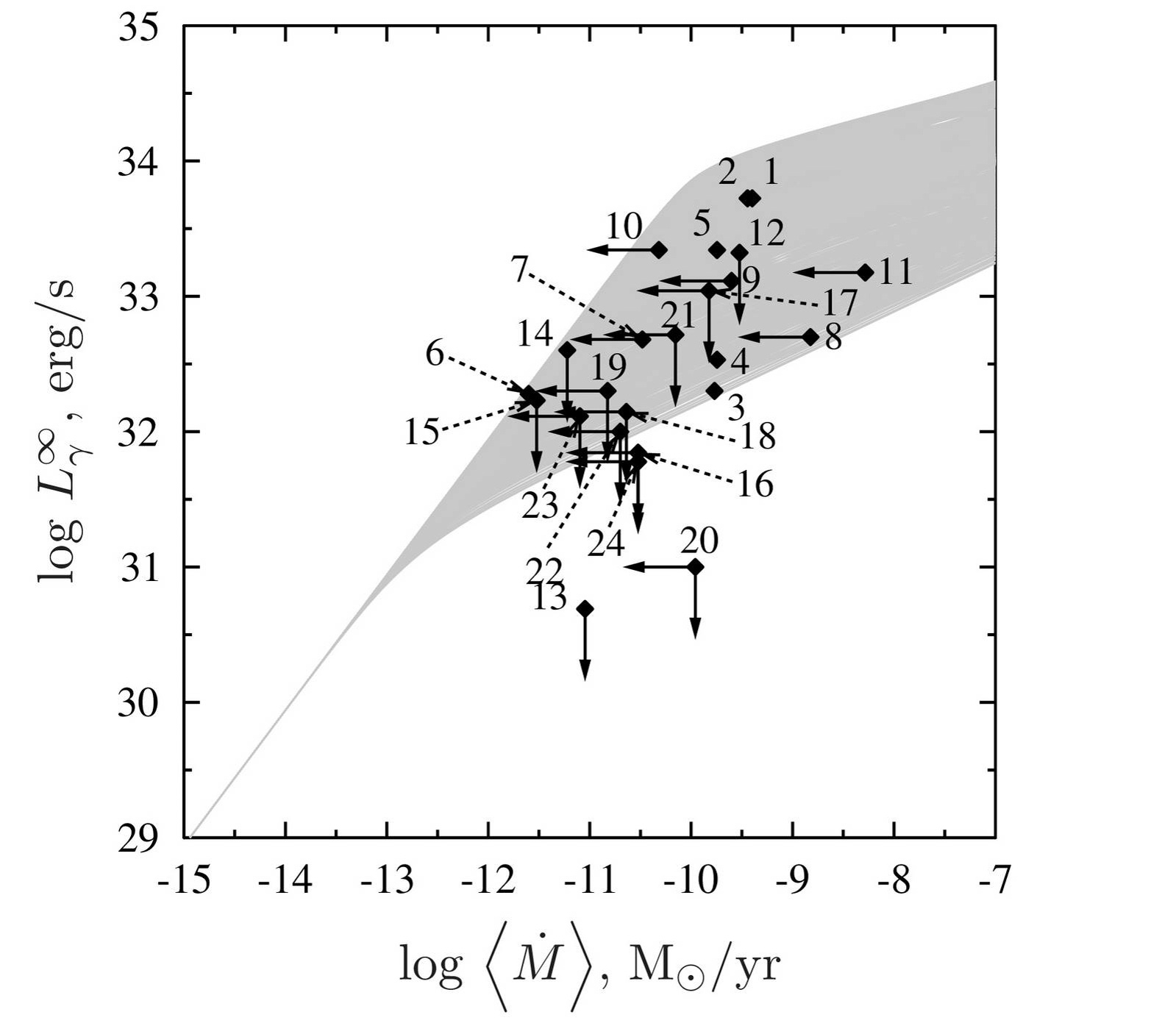}
\caption{Same as in Fig.\ \ref{fig:heatCR_100},  but the
neutrino emissivity is multiplied by a factor of  $\beta=10^{-4}$.
The two populations of stars merge, but the coldest sources (and some warmer ones too) contradict this model.}
\label{fig:heatCR_10000}
\end{figure}

Finally, let us assume a weakly enhanced neutrino emission with
 $\beta = 10^{-4}$ (Figs.\ \ref{fig:coolCR_10000} and \ref{fig:heatCR_10000}). If we take the heat blanketing envelopes made of iron and
do not broaden the threshold of the enhanced neutrino emission, we
would again obtain two distinct populations of rather warm ($M \leq
M_{\rm D}$) and less warm ($M>M_{\rm D}$) neutron stars separated by
a ``gap.'' If, however, we introduce the averaging over masses of light
elements in the heat blanketing envelopes, the two populations will
merge into a single one (almost no broadening of the threshold for the
enhanced neutrino emission is required!). Then we would be able to
explain the data on isolated cooling neutron stars. However, we
would be unable to interpret the observations of XRTs, especially
the coldest source SAX 1808.4--3658 (see Figs.\ \ref{fig:coolCR_10000} and \ref{fig:heatCR_10000}).

Therefore, our models of neutron stars whose neutrino cooling is
less enhanced than the direct Urca process can (in principle)
explain the data on isolated neutron stars but cannot explain the
data on quasi-stationary neutron stars in XRTs.

\section{Conclusions}
\label{s:concl}

We have proposed a statistical theory of thermal evolution of
cooling isolated middle-aged neutron stars and old transiently
accreting quasi-stationary neutron stars in XRTs. The theory is
based on the standard theory of neutron star cooling and heating
added by important elements of statistical theory such as mass
distributions of isolated and accreting neutron stars and mass
distributions of light elements in heat blanketing envelopes of
these stars. Instead of traditional cooling and heating curves we
introduce the probability to find cooling and heating neutron stars
in different parts of $T_{\rm s}^\infty-t$ and
$L_\gamma^\infty-\langle \dot{M} \rangle$ diagrams, respectively.
These probabilities have been compared with observations of
neutron stars of both types.

We have considered the simplest version of the statistical theory.
We have taken one EOS of nucleon matter in the neutron star core
($M_{\rm max}=2.16\,{\rm M_\odot}$) where the powerful direct Urca
process is switched on at $\rho>\rho_{\rm D}=1.00 \times 10^{15}$
g~cm$^{-3}$ ($M>M_{\rm D}=1.72\,{\rm M_\odot}$). We have introduced
phenomenologically the broadening of the direct Urca threshold,
distribution functions over neutron star masses (different for
isolated and transiently accreting neutron stars) and calculated the
required probabilities. We have varied the broadening of the direct
Urca threshold [the parameter $\alpha$ in equation
(\ref{e:broadD})], and typical mass ranges of isolated and accreting
neutron stars. In this way we have obtained a reasonable agreement
with observations of isolated and accreted neutron stars for
$\alpha=0.1$, $\mu_\mathrm{i}=1.4$, $\sigma_\mathrm{i}=0.15$,
$\mu_\mathrm{a}=0.47$, and $\sigma_\mathrm{i}=0.17$.

This explanation of \emph{all the data} essentially requires (i) the
presence of the direct Urca process in the inner cores of massive
neutron stars (to interpret the observations of SAX 1808.4--3658);
(ii) quite definite broadening of the direct Urca threshold
($\alpha\approx0.1$) to merge the populations of warm ($M\leq M_{\rm D}$)
and colder ($M>M_{\rm D}$) stars of each type into one (observable)
population; and (iii) higher typical masses of accreting stars (to
explain the very cold accreting source SAX 1808.4--3658 and the
absence of very cold middle-aged isolated neutron stars). In this
scenario the averaging over the masses of light elements in the heat
blanketing envelopes plays relatively minor role but is helpful to
explain the existence of warmer isolated and accreting sources.
Nevertheless, these sources can be explained by assuming the
presence of strong proton superfluidity in stars with $M<M_{\rm D}$.
This superfluidity suppresses the modified Urca process, which is
the major process of neutrino emission in low-mass stars. Such
stars will become slower neutrino coolers, and hence warmer sources
(e.g., \citealt{YP04}, and references therein). The
required broading of the direct Urca threshold can also be produced
by weakening of proton superfluidity in the massive stars (Section
\ref{s:broadUrca}). Therefore, the obtained explanation, in physical
terms, can be reached by assuming the presence of proton
superfluidity in neutron star cores. This superfluidity should be
strong in low-mass stars but weaken in high-mass ones whose neutrino
emission is greatly enhanced by the direct Urca process.

The present scenario is different from the minimal cooling model
\citep{GKYG04,PLPS04}. The latter model assumes
that the enhanced neutrino cooling is produced by the neutrino
emission due to the triplet-state pairing of neutrons. This
enhancement is much weaker than that due to the direct Urca process;
it cannot explain the observations of SAX 1808.4--3658.

On the other hand, recent analysis of the observations of the
neutron star in the Cassiopeia A (Cas A) supernova remnant by
\citet{HoHeinke_09} and \citet{HH10} indicated that this neutron
star has carbon atmosphere, is sufficiently warm but shows rather
rapid cooling in real time (with the surface temperature drop by a
few percent in about 10 years of observations). These results have
been explained \citep{PPLS11,Shternin_etal11} within the minimum
cooling model, by a  neutrino outburst within the star due to
moderately strong triplet-state pairing of neutrons. However, the
presence of real-time cooling has been put into question by
\citet{PPSK13} who attribute it to the Chandra ACIS-S detector
degradation in soft channels. A detailed analysis of the Cas~A
surface temperature decline has been done recently by
\citet{Elshamouty_etal13} by comparing the results from all the {\it
Chandra} detectors with the main conclusion that the real time
cooling is available although somewhat weaker than obtained before.
Thus the problem of real time cooling of the Cas A neutron star
remains open. If it is available it cannot be explained by the
scenario suggested in this paper.

Let us mention other results of this paper which seem original.
First, we have shown that if the direct Urca threshold is not
broadened, there are two different populations of accreting neutron
stars, warmer and colder ones, separated by a large ``gap.'' Second, we
have obtained that if the neutrino emission in massive stars is
enhanced only slightly ($\beta\sim 10^{-4}$, Section
\ref{s:pion-kaon}), then the averaging over different amounts of
light elements in the heat blanketing envelopes merges the populations
of warmer and colder (isolated and accreting) stars into one
population even without broadening the threshold of the enhanced
neutrino emission.

There is no doubt that the statistical theory presented above can be
elaborated further. For instance, we have used only one EOS of
superdense matter in neutron star cores, while one can try many
other ones. However, it is possible to predict that the results will
be similar, rescaled with respect to the values of $M_{\rm D}$ for
new EOSs. One can also try different mass distributions of isolated
and accreting neutron stars. In addition, one can expect that the
distribution of light elements in the heat blanketing envelopes is
not entirely arbitrary but is regulated by diffusion processes in
the envelopes. Another issue for future studies would be to include
the effects of rotation and magnetic fields, and also numerous
effects of nucleon superfluidity (see, e.g., Fig.\
\ref{fig:coolD_10}).

Note that statistical studies of populations of cooling neutron
stars have been performed in several publications (e.g., \citealt{PGTB06})
but under quite different approaches and with different conclusions.

\section*{acknowledgements}

The authors are grateful for the partial support by the State
Program ``Leading Scientific Schools of Russian Federation'' (grant
NSh 294.2014.2). The work of MB has also been partly supported by
the Dynasty Foundation, and the work of DY by Russian Foundation for
Basic Research (grants Nos. 14-02-00868-a and 13-02-12017-ofi-M) and
by ``NewCompStar,'' COST Action MP1304. In addition, DY
acknowledges sponsorship of the ISSI (Bern, Switzerland) within the
program ``Probing Deep into the Neutron Star Crust with Transient
Neutron-Star Low-Mass X-Ray Binaries.''


\end{document}